\documentclass{iopart}

\usepackage[utf8]{inputenc} 
\usepackage{iopams}
\expandafter\let\csname equation*\endcsname\relax
\expandafter\let\csname endequation*\endcsname\relax

\usepackage{mathtools}
\usepackage{amsmath}
\usepackage{amssymb}
\usepackage{amsthm}

\usepackage{graphicx}
\graphicspath{{./figs/}}
\usepackage[font={small}]{caption}
\usepackage{mwe}
\usepackage{dsfont}
\usepackage{bbm}
\usepackage{cite}

\newcommand{\ui}{\mathrm{i}}   
\newcommand{\ue}{\mathrm{e}}    
  
\newcommand{\Tp}{\textnormal{T}}
\newcommand{\defeq}{\vcentcolon=}
\newcommand{\pa}{\partial}
\renewcommand\Re{\operatorname{Re}}
\renewcommand\Im{\operatorname{Im}}
\newcommand{\la}{\langle}
\newcommand{\ra}{\rangle}

\newcommand{\R}{\mathbb{R}}
\newcommand{\C}{\mathbb{C}}

\begin{document}

\title[Lindblad dynamics of Gaussian states in the semiclassical limit]{Lindblad dynamics of Gaussian states and their superpositions in the semiclassical limit}
\author{E M Graefe$^1$, B Longstaff$^1$, T Plastow$^2$, and R Schubert$^2$}
\address{$^1$Department of Mathematics, Imperial College London, London SW7 2AZ, United Kingdom
\\
$^2$ Department of Mathematics, University of Bristol, Bristol, BS8 1TW, United Kingdom}
\begin{abstract}
The time evolution of the Wigner function for Gaussian states generated by Lindblad quantum dynamics is investigated in the semiclassical limit. A new type of phase-space dynamics is obtained for the centre of a Gaussian Wigner function, where the Lindblad terms generally introduce a non-Hamiltonian flow. In addition to this, the Gaussian approximation yields dynamical equations for the covariances. The approximation becomes exact for linear Lindblad operators and a quadratic Hamiltonian. By viewing the Wigner function as a wave function on a coordinate space of doubled dimension, and the phase-space Lindblad equation as a Schr\"{o}dinger equation with a non-Hermitian Hamiltonian, a further set of semiclassical equations are derived. These are capable of describing the interference terms in Wigner functions arising in superpositions of Gaussian states, as demonstrated for a cat state in an anharmonic oscillator subject to damping.  
\end{abstract}

\section{Introduction}
Since the early days of quantum mechanics the dynamics of Gaussian wave packets has been considered as a natural connection between quantum and classical dynamics \cite{Schr26,Dira58}. As shown by Heller, Hepp and Littlejohn \cite{Hell75,Hepp74,Litt86}, in closed quantum systems the motion of the centre in the semiclassical limit is described by Hamilton's equations of motion. Semiclassical methods based on the time evolution of Gaussian states along classical trajectories provide powerful numerical and analytical tools \cite{Hell81,Hube88b,Hube89,Mill02,Bara01,Kong16}. 

Realistic quantum systems, however, are open. That is, they exchange energy with their environment. In the Markovian approximation a system weakly coupled to its environment can be described by a Lindblad equation (see, e.g. \cite{Breu02}). Markovian open quantum systems play a crucial role in various branches of quantum physics ranging from quantum optics and information to atomic, nuclear, and condensed matter physics. It is an interesting question how the dynamics of Gaussian wave packets generalise in this context, and what can be learned from the semiclassical limit. Here we generalise the approach of Heller and Littlejohn \cite{Hell75,Litt86} to Lindblad type quantum dynamics. This yields a new type of phase-space dynamics for the position and momentum expectation values, where the Lindblad terms generally introduce non-Hamiltonian flows, which in special cases can take the form of a gradient flow. Furthermore, the semiclassical dynamics yields an approximation of the quantum covariances, which can be useful in various applications. 

The current study complements previous investigations of the semiclassical limit of Lindblad dynamics, using the framework of path integrals \cite{Stru97}, and complex WKB dynamics \cite{Ozor09}. In \cite{Brod10b} first steps have been made towards applying Heller's wave packet method to Lindblad dynamics for the case of linear Lindblad operators. There the framework of chord functions in a doubled phase space is used. This differs from the approach taken here, which is a more direct application of Heller's method.  
 
The paper is organised as follows. We first introduce the Lindblad equation on phase space in section \ref{sec_Lindblad_PS} and make a Gaussian ansatz for the state, from which semiclassical equations of motion for the centre and covariance matrix are derived in section \ref{sec_Gaussian}. The resulting semiclassical equations are interpreted and the conditions under which the centre dynamics may be written as a gradient flow are discussed. We transform our equations of motion to a form that is better suited to many applications, particularly those in quantum optics. Two examples are used to illustrate this. An oscillator with nonlinear losses and amplification, and the dissipative Bose-Hubbard Hamiltonian. Finally, by viewing the phase-space Lindblad equation as a Schr\"{o}dinger equation with a non-Hermitian Hamiltonian, we connect the Lindblad dynamics to non-Hermitian quantum dynamics in section \ref{section:non herm}. The semiclassical limit of the latter has been investigated in detail by two of the authors in \cite{Grae11,Grae12}. Applying results from this context allows us to derive a further set of semiclassical equations for Lindblad dynamics. These are capable of describing the interference terms in Wigner functions arising, for example, in superpositions of Gaussian states. We demonstrate this for a cat state in a damped anharmonic oscillator. We conclude with a short summary. 

\section{Lindblad Equation on Phase Space}
\label{sec_Lindblad_PS}
We consider the dynamics of quantum systems generated by equations of Lindblad type
\begin{equation}
\label{eqn:lindblad-eq}
i\hbar \frac{\pa \hat{\rho}}{\pa t} = [\hat{H},\hat{\rho}] + i\sum_k \hat{L}_k \hat{\rho} \hat{L}_k^\dagger - \frac{1}{2}\hat{L}_k^\dagger \hat{L}_k \hat{\rho} - \frac{1}{2}\hat{\rho} \hat{L}_k^\dagger \hat{L}_k,
\end{equation}
where $\hat\rho$ is the density matrix describing the state of the quantum system. The first term corresponds to the unitary dynamics generated by the Hamiltonian $\hat{H}$, while the following terms containing the Lindblad operators $\hat{L}_k$ account for (weak) interactions of the system with an environment. For more details see, e.g., \cite{Breu02}.

The phase-space representation of quantum mechanics \cite{Zach05} is particularly convenient for analysing the semiclassical limit. In the Wigner-Weyl representation an operator $\hat f$ on Hilbert space is mapped to a phase-space function $f(x)$, known as the Weyl symbol of $\hat f$, via the Wigner-Weyl transformation
\begin{equation}
\label{eqn:wigner weyl}
f(x) = \int d\xi \la q+\frac{\xi}{2}| \hat f | q - \frac{\xi}{2} \ra \ue^{-ip\cdot \xi/\hbar},
\end{equation}
where the canonical coordinates $x=(q,p)$ span the $2n$-dimensional phase space. In this representation the quantum state $\hat \rho$ is represented by the Wigner function, which is defined by the Wigner-Weyl transformation of the operator $\hat \rho$ as
\begin{equation}
\label{eqn:wigner func}
W(x) = \frac{1}{(2\pi\hbar)^n}\int d\xi \la q+ \frac{\xi}{2}|\hat \rho |q - \frac{\xi}{2} \ra \ue^{-ip\cdot \xi/\hbar}.
\end{equation}
Assuming that the Hamiltonian $\hat{H}$ and the Lindblad operators $\hat{L}_k$ are the Weyl quantisations of sufficiently well-behaved phase-space functions $H(x)$ and $L_k(x)$, the evolution equation for the Wigner function is given by
\begin{equation}
\label{eqn:wigner moyal eom}
i\hbar \frac{\pa W}{\pa t} = H\star W-W\star H+i\sum_k L_k\star W\star\bar{L}_k - \frac{1}{2} \bar{L}_k\star L_k\star W - \frac{1}{2}W\star \bar{L}_k\star L_k,
\end{equation}
where $f \star g$ denotes the Moyal product of two phase-space functions $f$ and $g$,
\begin{align}
(f\star g)(x)&=f(x)\ue^{\frac{i\hbar}{2} \overleftarrow{\nabla}\cdot\Omega\overrightarrow{\nabla}}g(x), \nonumber \\ &= f(x)g(x) + \frac{i\hbar}{2}\{f(x),g(x)\} + \ldots. \label{eqn:Moyal product1}
\end{align}
Here $\Omega$ denotes the symplectic form
\begin{equation}
\label{eqn:symplectic}
\Omega = \begin{pmatrix} 0 & I_n \\ -I_n & 0\end{pmatrix},
\end{equation}
the phase-space gradient is $\nabla\defeq\left(\pa_q,\pa_p\right)$ and the arrows over the differential operators in (\ref{eqn:Moyal product1}) indicate whether they act on the function to the left or to the right. The first two terms of the Moyal product are shown in  (\ref{eqn:Moyal product1}), where  $\{A,B\}=\nabla A \cdot \Omega\nabla B$ is the usual Poisson bracket.

Following the work of Heller, Hepp and Littlejohn \cite{Hell75,Hepp74,Litt86} for closed systems, assuming that the initial state is a well-localised Gaussian state, it is justified to approximate the Hamiltonian and the Lindblad operators by finite Taylor expansions around the centre of the state. We make the ansatz that the time-evolved state remains Gaussian for all times and instantaneously expand the Hamiltonian and the Lindblad operators in Taylor series around the time-dependent centre of the state. For closed quantum systems in leading orders of $\hbar$ this approximation yields Hamilton's canonical equations of motion for the centre of the state, while the width changes with the linearised classical flow. In what follows we shall investigate how this dynamics generalises in the presence of Lindblad operators. 

\section{Gaussian Evolution in the Semiclassical Limit} \label{gaussian evo sec}
\label{sec_Gaussian}
\subsection{Deriving semiclassical equations of motion}
The Wigner function of a general Gaussian state is of the form
\begin{equation}
W(x)=\frac{\sqrt{\det{G}}}{(\pi\hbar)^n}\ue^{-(x-X)\cdot G(x-X)/\hbar},
\label{eqn:wigner gaussian}
\end{equation}
where $x=(q,p) \in \mathbb{R}^n \times \mathbb{R}^n$ are canonical phase-space coordinates, $X$ is a vector of the expectation values of the quantum position and momentum operators 
\begin{equation}
X_k = \la \hat{x}_k \ra \equiv \textrm{Tr}(\hat\rho \hat{x}_k),
\label{eqn:first moments}
\end{equation}
with $\hat{x}=(\hat{q}_1,\ldots,\hat{q}_n,\hat{p}_1,\ldots,\hat{p}_n)$, and $G$ is a real, symmetric and positive definite matrix, encoding the width of the wave packet via the (co)variances of the canonical operators as
\begin{equation}
\hbar\left(G^{-1}\right)_{jk} = \langle \hat x_j \hat x_k + \hat x_k \hat x_j \rangle-2\langle \hat x_j \rangle\langle \hat x_k\rangle.
\end{equation}

An initial Gaussian state, with time-dependent parameters $G$ and $X_k$, remains Gaussian for all times under the quantum dynamics generated by (\ref{eqn:wigner moyal eom}) if the Hamiltonian is at most quadratic and the Lindblad operators are linear in $\hat x$, \cite{Brod10b}. We want to show that to leading order in $\hbar$  this remains  the case for general Hamiltonians and Lindblad operators. To this end we first have to recall how 
the semiclassical expansion (\ref{eqn:Moyal product1}) scales with $\hbar$ if we insert a Gaussian (\ref{eqn:wigner gaussian}) that is localised in phase space on a scale of $\sqrt\hbar$. If we rewrite $W(x)=\hbar^{-n} w_0(\hbar^{-1/2}(x-X))$ with $w_0(x)=\frac{\sqrt{\det G}}{\pi^n}\exp(x\cdot Gx)$, then we find for arbitrary derivatives 
\begin{equation}\label{eq:der-W}
\pa^{\alpha}_xW(x)=\hbar^{-|\alpha|/2}  \hbar^{-n} w_0^{(\alpha)}(\hbar^{-1/2}(x-X))\, \quad \text{with}\quad w_0^{(\alpha)}(y)=\pa_y^{\alpha}w_0(y)\,\, .
\end{equation}
Hence, every derivative contributes a factor $1/\sqrt{\hbar}$ and the semiclassical expansion of the Moyal bracket \eqref{eqn:Moyal product1} effectively becomes an expansion in powers of 
$\sqrt{\hbar}$ if $g=W$.  Furthermore, if we encounter terms of the form $A(x) W(x)$ we can Taylor expand $A$ around $x=X$ and use that for any multi-index $\beta$ 
\begin{equation}\label{eq:pow-W}
(x-X)^{\beta}W(x)=\hbar^{|\beta|/2} \hbar^{-n} w_{\beta}(\hbar^{-1/2}(x-X))\, \quad \text{with}\quad w_{\beta}(y)=y^{\beta}w_0(y)\,\, .
\end{equation}
Thus, multiplying by powers of $(x-X)$ reduces the size by powers of $\sqrt{\hbar}$ so that higher order terms in the Taylor expansion contribute to higher orders in semiclassical expansions in powers of $\sqrt{\hbar}$. 

We now expand (\ref{eqn:Moyal product1}) in the evolution equation for the Wigner function (\ref{eqn:wigner moyal eom}) and by \eqref{eq:der-W} we only need to take the first two terms into account, yielding
\begin{align}
\label{eqn:sc lindblad 1}
i\hbar \frac{\pa W}{\pa t}=i\hbar \{H,W\} &+ \ui\hbar\sum_k \Im\left(L_k\{\bar{L}_k, W\} \right)\nonumber \\&+\hbar\sum_k \{\bar{L}_k,L_k\} W+ \frac{i\hbar^2}{2}\sum_k\Re\left(\{L_k,\{\bar{L}_k,W\}\}\right)+\cdots ,
\end{align}
where the remaining terms are of order $\hbar^{3/2}$ or higher relative to $W$.  Note that the first two terms on the right-hand side are of order $\sqrt{\hbar}$ whereas the two remaining terms are of order $\hbar$. 
In the next step we expand $H$ and $L_k$ around the centre $X$, i.e., in powers of $\delta x:=x-X$, and by \eqref{eq:pow-W}  we need to expand $H$ and $L_k$ up to second order in the first two terms on the right-hand side, and $L_k$ up to first order in the second two terms in order to include all terms up to order $\hbar^{1}$ in (\ref{eqn:sc lindblad 1}). 
 
The first two terms on the right-hand side of (\ref{eqn:sc lindblad 1}) then become 
\begin{equation}
2iG \bigg[\Omega \nabla H +\Omega\sum_k\Im (L_k\nabla \bar{L}_k)\bigg]\cdot \delta x\, W -2i \delta x\cdot \Lambda \Omega G \delta x\,W+\cdots 
\end{equation}
with 
\begin{equation}
\Lambda:=H''+\sum_k \Im \big(L_k \bar{L}_k''\big)+\sum_k\Im\big(\nabla L_k\nabla \bar{L}_k^T\big)\,\, .
\end{equation}
For the final two terms on the right-hand side of  (\ref{eqn:sc lindblad 1}) we only need to expand the Lindblad terms up to first order and obtain 
\begin{equation}
\hbar \sum_k \big[\nabla \bar{L}_k\cdot \Omega \nabla L_k+\nabla L_k\cdot \Omega G\Omega \nabla \bar{L}_k\big]\, W
-2\delta x\cdot G\Omega D\Omega G\delta x\, W+\cdots 
\end{equation}
where 
\begin{equation}
D=\sum_k\Re\big(\nabla L_k \nabla \bar{L}_k^T\big)\,\, .
\end{equation}
Finally, using 
\begin{equation}
\ui\hbar \pa_t W=\bigg[\frac{i\hbar}{2}\Tr(G^{-1}\dot G) + 2iG\dot X \cdot \delta x - i\dot G \delta x \cdot \delta x\bigg]W
\end{equation}
we can separate different powers of $\delta x$ in \eqref{eqn:sc lindblad 1}, which leads to the semiclassical equations of motion for the parameters $X$ and $G$
\begin{align}
\dot X &= \Omega\nabla H + \Omega \sum_k \Im \left(L_k \nabla\bar{L}_k\right), \label{eqn:X eom} \\
\dot{G} &= \Lambda\Omega G - G \Omega \Lambda^T+ 2 G \Omega D\Omega G
\label{eqn:G eom}.
\end{align}
To obtain (\ref{eqn:G eom}) the symmetry enforcing convention $G = (G+G^\Tp)/2$ was applied. This is not an approximation and is done purely for cosmetic reasons; as the Wigner function depends only on the symmetric part of $G$, any antisymmetric part is unobservable. There is a third equation for $\Tr G^{-1}\dot G$ but this is automatically satisfied if (\ref{eqn:G eom}) holds.


Equations (\ref{eqn:X eom}) and (\ref{eqn:G eom}) are two of the main results of the present paper. As expected, in the unitary case the centre moves according to the classical canonical equations of motion $\dot X=\Omega\nabla H$ and the evolution of $G$ is governed by the linearised Hamiltonian flow around the classical trajectory. The general dynamical equations for the centre of the state can be interpreted as a generalisation of Hamilton's canonical equations to Markovian open quantum systems. Following Littlejohn \cite{Litt86}, the terms $\Lambda \Omega G-G \Omega \Lambda^T$ can be interpreted as a linearised flow in the non-unitary case. However, the additional term $2G \Omega D\Omega G$ in (\ref{eqn:G eom}) does not result from the linearised flow. This term originates from the double Poisson brackets in (\ref{eqn:sc lindblad 1}) and ensures that the Heisenberg uncertainty principle is not violated. In particular, a physically meaningful Gaussian state must fulfil the Robertson-Schr\"{o}dinger uncertainty relation, expressed in terms of $G$ as \cite{Ades14}
\begin{equation}
\label{eqn:Heisenberg}
G^{-1} + i \Omega \geq 0.
\end{equation}
The extra term $2G \Omega D\Omega G$ is a quantum correction appearing in the semiclassical dynamics, guaranteeing that (\ref{eqn:Heisenberg}) is fulfilled for all times. 

The generalisation of Heller's theory to Markovian open systems with \textit{linear} Lindblad operators has previously been considered in \cite{Ozor09,Brod10b}. The approach taken therein, however, is quite different from the one presented here. In particular, the authors work with the Fourier transform of the Wigner function, the so-called chord function, in double phase space. While an equivalent of equations (\ref{eqn:X eom}) and (\ref{eqn:G eom}) is implicitly contained in the results presented there, they are not explicitly derived or analysed.  Of course, both approaches are exact for quadratic Hamiltonians and linear Lindblad operators, and, as we shall discuss in detail in section \ref{section:non herm}, the equations in \cite{Ozor09,Brod10b} can be shown to reduce to a special case of our equations (\ref{eqn:X eom}) and (\ref{eqn:G eom}) for linear Lindblad operators. Apart from going beyond the restriction of linear Lindblad operators, the approach presented here has the advantage of directly yielding the set of dynamical equations (\ref{eqn:X eom}) and (\ref{eqn:G eom}) for the position and momentum expectation values as well as the covariances.

\subsection{Geometric interpretation of the Lindblad terms}
We now analyse the structure of the Lindblad terms in the dynamical equation for the centre (\ref{eqn:X eom}) in more detail. There are two cases for which the geometric interpretation of these terms is simple: First, for purely Hermitian or anti-Hermitian Lindblad operators the flow generated by the Lindblad terms vanishes, and the only effect in the semiclassical description is on the width of the Wigner function. This is in line with the well-known result that purely Hermitian or anti-Hermitian Lindblad operators lead to decoherence but no dissipation \cite{Perc98}. 

The second case is where the Weyl symbols of the Lindblad operators are holomorphic functions of $q\pm ip$. In this case, the flow they generate can be written as a gradient flow of the phase-space function $\Gamma\defeq \mp \frac{1}{2}|L|^2$. In fact, the flow generated by a Lindblad operator is the gradient flow of the function $\Gamma=\mp\frac{1}{2}|L|^2$ if and only if the Lindblad symbol is a holomorphic function of either $q+i p$ or $q-i p$. This can be seen as follows. The holomorphy of $L$ as a function of $q \pm i p$ is equivalent to the validity of the Cauchy-Riemann conditions 
\begin{equation}
\nabla \Re(L) = \pm\Omega \nabla \Im(L), 
\end{equation}
which imply 
\begin{equation}
\nabla \Im(L) = \mp\Omega \nabla \Re(L). 
\end{equation}
Observing that the Lindblad term on the right hand side of (\ref{eqn:X eom}) may be expressed as
\begin{equation}
\label{eqn:X eom 2}
\Omega \Im \left(L \nabla\bar{L}\right) = \Im (L) \Omega \nabla \Re (L) - \Re (L) \Omega \nabla \Im (L), 
\end{equation}
the right hand side can immediately be identified as
\begin{equation}
\label{eqn:grad L2}
\mp(\Re(L) \nabla \Re(L) + \Im(L) \nabla \Im(L)) = \mp\frac{1}{2} \nabla |L|^2.
\end{equation}
On the other hand, assume we have 
\begin{equation}
\label{eqn:assump}
\Im(L) \Omega \nabla \Re(L) - \Re(L) \Omega \nabla \Im(L) =  \mp\frac{1}{2} \nabla |L|^2 = \mp\big(\Re(L) \nabla \Re(L) + \Im(L) \nabla \Im(L)\big)
\end{equation}
and that $\Re(L) \neq 0$ and $\Im(L) \neq 0$ (as otherwise the Lindblad flow vanishes). Acting with the symplectic onto equation (\ref{eqn:assump}) gives
\begin{equation}
-\Im(L) \nabla \Re(L) + \Re(L) \nabla \Im(L) =  \mp(\Re(L)\Omega  \nabla \Re(L) + \Im(L) \Omega\nabla \Im(L)).
\end{equation}
Combining this with the original expression  (\ref{eqn:assump}) yields the Cauchy-Riemann conditions $\nabla \Re(L) = \pm\Omega \nabla \Im(L)$. The gradient dynamics drives trajectories towards the closest maximum of the function $\Gamma$. Thus one possible physical interpretation is that $\Gamma$ is an entropy of the system. Note that similar gradient dynamics have been discussed in the context of thermodynamics e.g. by \"Ottinger and Grmela \cite{Grme97,Oett97}. 

Of course, there are operators that are neither Hermitian or anti-Hermitian nor have symbols that are holomorphic functions of $q\pm i p$. In some of these cases the Lindblad terms can still be written as gradient flows of more general functions. For instance, Lindblad operators of the form $\hat L=a\hat q+i b \hat p$ with $a\neq \pm b$ result in a semiclassical flow given by the gradient of the function $\Gamma=-\frac{ab}{2}(q^2+p^2)\neq \mp\frac{1}{2}|L|^2$. 

There are other Lindblad operators that instead lead to Hamiltonian flows, such as normal operators in one dimension. However, in general the Lindblad term leads to a flow which is neither a Hamiltonian nor a gradient flow. Instead of deriving more intricate mathematical conditions for different types of flows, the following two examples aim to develop a better intuition of Lindblad operators that do not generate a Hamiltonian or gradient flow.

As an example, consider a two-dimensional system and a linear Lindblad operator $\hat L = \sqrt{\gamma}(\hat q_1 + i\hat p_2)$. Without a Hamiltonian term the evolution equations for the centre are 
\begin{equation}
\label{eqn:no H}
\dot{q}_1 = 0, \qquad
\dot{q}_2 = -\gamma q_1, \qquad
\dot{p}_1 = -\gamma p_2,\qquad
\dot{p}_2 = 0.
\end{equation}
In this case the semiclassical approximation is exact since $L$ is linear. Both $q_1$ and $p_2$ are constants of motion and $p_1(q_2)=\frac{p_2}{q_1}q_2+\frac{p_1(0)}{q_2(0)}$, i.e. all straight lines are phase-space trajectories and all points in the plane $(0,q_2,p_1,0)$ in the four-dimensional phase space are fixed points.

\begin{figure}[htb]
      \centering
           \includegraphics[width=0.5\textwidth]{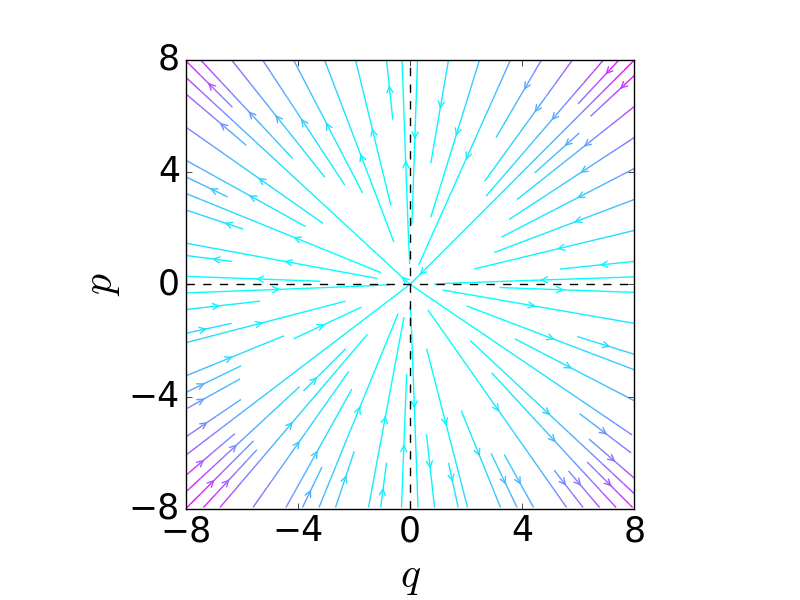}
\caption{The phase-space portrait for the classical dynamics (\ref{eqn:X ngf}) with $\gamma=0.1$. False colours indicate the velocity.} 
\label{fig:non-grad1}
\end{figure}

A less trivial example is a one-dimensional system with nonlinear Lindblad operator $\hat L = \sqrt{\gamma}(\hat q^2 + i \hat p^2)$. The semiclassical equations of motion for $q,p$ are found to be
\begin{align}
\label{eqn:X ngf}
\dot q &= - 2\gamma q^2 p, \\
\dot p &= - 2\gamma q p^2. \nonumber
\end{align}
The phase-space portrait of this dynamics is depicted in Figure \ref{fig:non-grad1}. The flow conserves the quantity $p/q$ and thus the trajectories are straight lines. The lines $q=0$ and $p=0$ are fixed points. In particular the point $p=0=q$ acts as a hyperbolic fixed point. Futhermore, the lines $q=p$ and $q=-p$ are the stable and unstable manifolds, respectively.

In this example the semiclassical description is of course only an approximation. We shall not dwell on its quantum-classical correspondence here but instead consider more physically relevant examples in detail later.

\subsection{Formulation in creation and annihilation operators}
In many applications, in particular in quantum optics, the Hamiltonian and Lindblad operators are expressed in terms of annihilation and creation operators $\hat{a}_j$ and $\hat{a}_j^\dagger$, satisfying the commutation relations
\begin{equation}
\label{eqn:boson com}
[\hat{a}_i,\hat{a}_j^\dagger]=\delta_{ij}, \quad [\hat{a}_i,\hat{a}_j]=0.
\end{equation} 
For such systems it is convenient to express the semiclassical equations (\ref{eqn:X eom}) and (\ref{eqn:G eom}) in terms of the complex canonical phase-space variables 
\begin{equation}
\label{eqn:complex phase var}
a_j = \frac{1}{\sqrt{2}}(q_j + ip_j),
\end{equation}
where $q_j$ and $p_j$ are the classical counterparts of the quadrature operators, defined as
\begin{equation}
\hat{q}_j=\frac{1}{\sqrt{2}}(\hat{a}_j^\dagger + \hat{a}_j),\quad \hat{p}_j=\frac{i}{\sqrt{2}}(\hat{a}_j^\dagger - \hat{a}_j),
\label{eqn:quad def}
\end{equation}
which are in general not associated to a physical position or momentum. For the remainder of this subsection we choose to work in units of $\hbar = 1$ for simplicity.

The operators $\hat{q}_j,\hat{p}_j$ can be grouped into the vector $\hat{x}=(\hat{q}_1,\ldots,\hat{q}_n,\hat{p}_1,\ldots,\hat{p}_n)$ and the commutation relations between the quadrature operators can then be written in the form
\begin{equation}
\label{eqn:qp comm}
[\hat{x}_i,\hat{x}_j] = i \Omega_{ij},
\end{equation}
where $\Omega$ is the symplectic form (\ref{eqn:symplectic}). That is, they form a set of canonically conjugate observables. If the Hamiltonian and Lindblad operators are expressed in terms of the $\hat{x}_j$, then the semiclassical Gaussian dynamics derived in the previous section are applicable. The transformation from quadrature operators $\hat{q}, \hat{p}$ to mode operators $\hat{a}, \hat{a}^\dagger$ is achieved via the transformation matrix
\begin{equation}
\label{eqn:T matrix}
T = \frac{1}{\sqrt{2}}\begin{pmatrix} I_n & iI_n\\ I_n & -iI_n\end{pmatrix}.
\end{equation}
Applying the transformations $X_c = T X$ and $G_c = T G T^\dag$ yields
\begin{align}
\dot{X}_c &= -i\Omega \nabla H + \frac{1}{2} \Omega \sum_k \left(\bar{L}_k \nabla L_k - L_k \nabla \bar{L}_k \right), \label{eqn:Xc eom}\\
\dot{G}_c &= G_c  \Omega (K- \Gamma) - (\bar{K} + \bar{\Gamma}) \Omega G_c + G_c  \Omega \Xi \Omega G_c, \label{eqn:Sigc eom}
\end{align}
where $X_c = (a,\bar{a})$ and $\nabla \defeq \left(\pa_a,\pa_{\bar{a}}\right)$. We have also defined $K$, $\Gamma$ and $\Xi$ as 
\begin{equation}
\label{eqn:K matrix}
K = iH'' + \frac{1}{2}\sum_k L_k \bar{L}_k''  - \bar{L}_k L_k'',
\end{equation}
\begin{equation}
\label{eqn:complex lambda}
\Gamma = \frac{1}{2}\sum_k \left(\nabla L_k \nabla \bar{L}_k^T - \nabla \bar{L}_k \nabla L_k^T \right),
\end{equation}
\begin{equation}
\label{eqn:complex D}
\Xi = \sum_k \left(\nabla L_k \nabla \bar{L}_k^T + \nabla \bar{L}_k \nabla L_k^T \right) \begin{pmatrix} 0 & I_n \\ I_n & 0 \end{pmatrix},
\end{equation}
where primes indicate Hessian matrices. Care must be taken when using these expressions.  For instance, the term $\bar{L}_k''$ in (\ref{eqn:K matrix}) is the Hessian matrix of $\bar{L}_k$ with respect to $(a,\bar a)$ and not the complex conjugate of the Hessian $L_k''$. The same applies to the gradients in (\ref{eqn:complex lambda}) and (\ref{eqn:complex D}). 

In the complex coordinates $(a,\bar{a})$ the covariance matrix $\Sigma = G_c^{-1}$ takes the block form
\begin{equation}
\label{eqn:Sig block}
\Sigma = \begin{pmatrix} \bar{\alpha} & \bar{\beta} \\ \beta & \alpha \end{pmatrix},
\end{equation}
where
\begin{align}
\alpha_{ij} &= \la \hat{a}_i^\dag \hat{a}_j + \hat{a}_j \hat{a}_i^\dag \ra - 2 \la \hat{a}_i^\dag \ra \la \hat{a}_j \ra, \label{eqn:alpha} \\ \beta_{ij} &= \la \hat{a}_i^\dag \hat{a}_j^\dag + \hat{a}_j^\dag \hat{a}_i^\dag \ra - 2 \la \hat{a}_i^\dag \ra \la \hat{a}_j^\dag \ra.
\end{align}

\subsection{Examples and applications}
\begin{figure}[htb]
      \centering
           \includegraphics[width=0.5\textwidth]{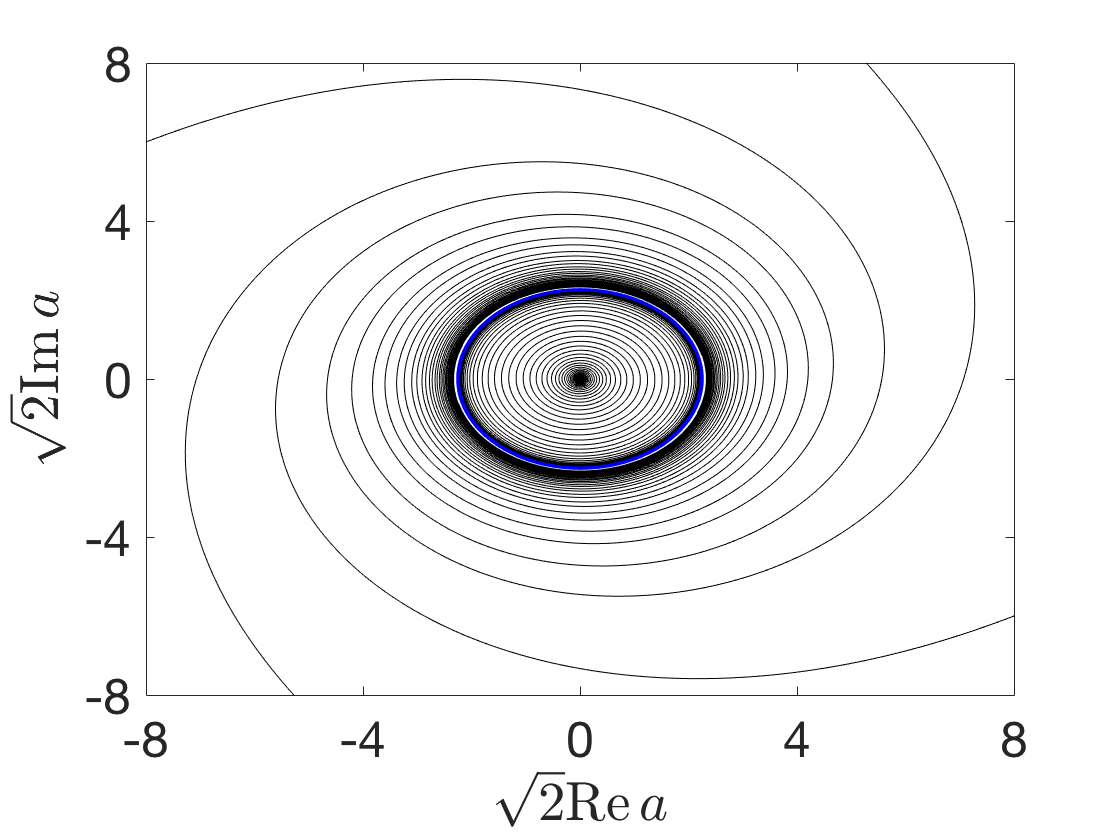}
\caption{A phase-space portrait for the classical dynamics (\ref{eqn:limit centre}) with $\omega=1$, $\gamma_1=0.1$, $\gamma_2=0.01$ and $A=0.15$. There is a stable limit cycle at $|a|^2 = 2.5$, indicated by the blue circle, and the origin is an unstable fixed point.}
\label{fig:limit pp}
\end{figure}

As a first example let us consider a harmonic oscillator with nonlinear damping and amplification. The Hamiltonian is given by
\begin{equation}
\hat H = \omega \hat{a}^\dag \hat a
\end{equation}
and the three Lindblad operators 
\begin{equation}
\hat{L}_1 = \sqrt{\gamma_1}\hat a,\quad \hat{L}_2 = \sqrt{\gamma_2}\hat a^2
\quad\, {\rm and}\quad \hat{L}_3 = \sqrt{A}\hat a^\dag
\end{equation}
describe the damping and gain, where $\gamma_1$, $\gamma_2$ and $A$ are the linear damping rate, nonlinear damping rate and amplification rate respectively. This model appears in the context of quantum optomechanics and is discussed in \cite{Bowe16}. For instance, it could describe a driven nanomechanical oscillator coupled to a thermal bath, where the damping rate depends on the excitation of the resonator. Applying (\ref{eqn:Xc eom}) immediately yields the semiclassical equation
\begin{equation}
\label{eqn:limit centre}
\dot a = -iwa + \frac{1}{2}(A-\gamma_1)a - \gamma_2|a|^2 a.
\end{equation}

 \begin{figure}[htb]
        \centering
            \includegraphics[width=0.32\textwidth]{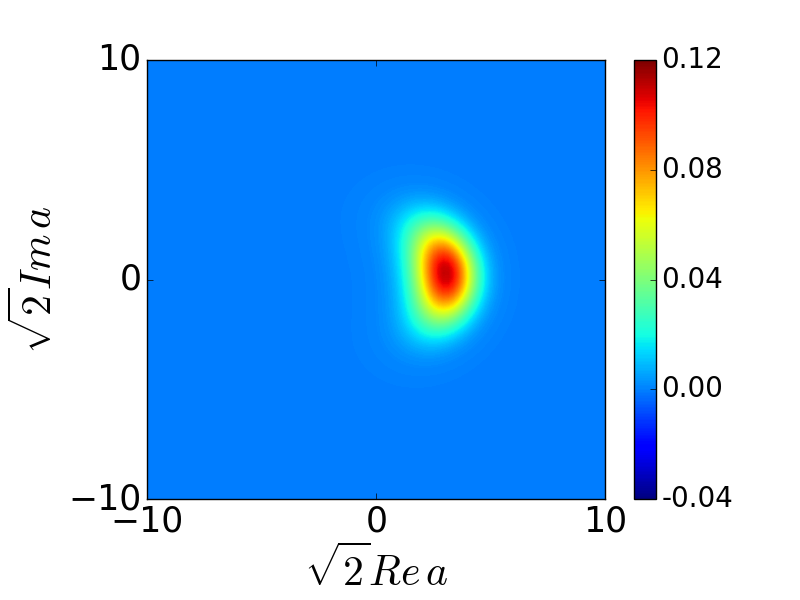}  
             \includegraphics[width=0.32\textwidth]{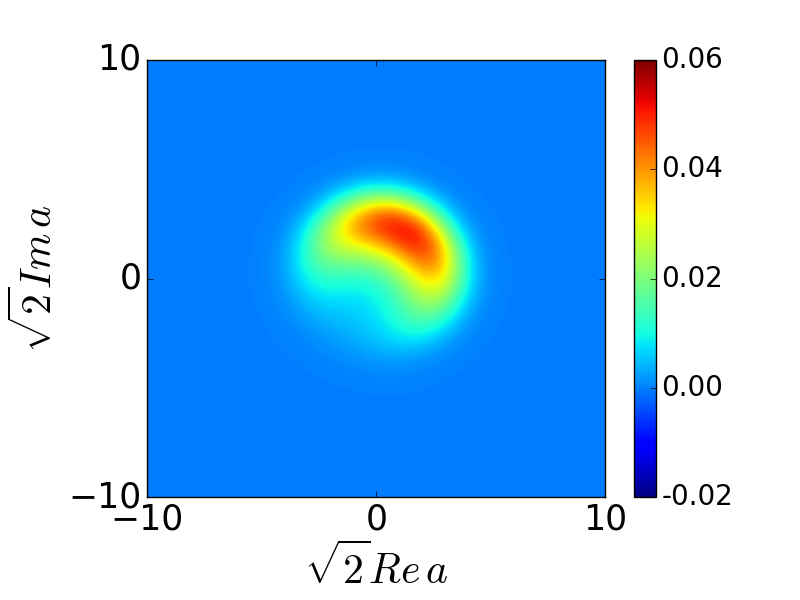}  
             \includegraphics[width=0.32\textwidth]{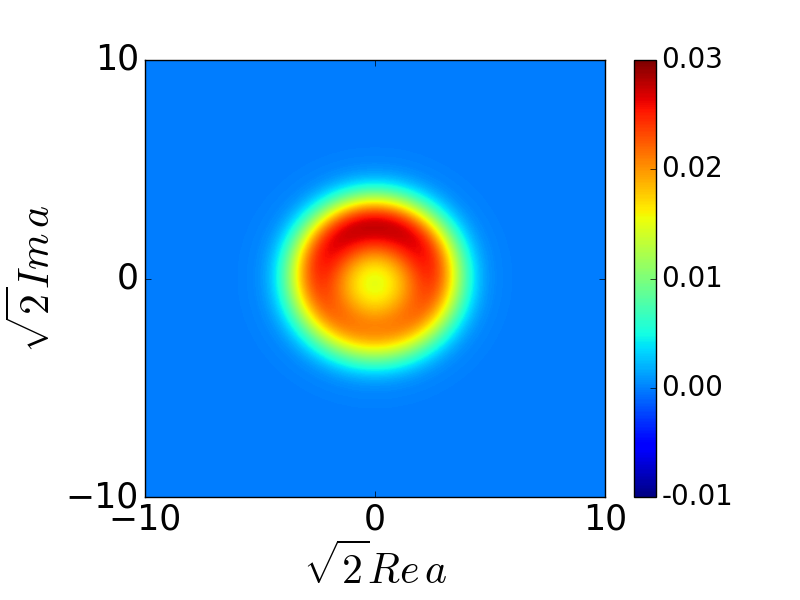}\hfill
             \includegraphics[width=0.32\textwidth]{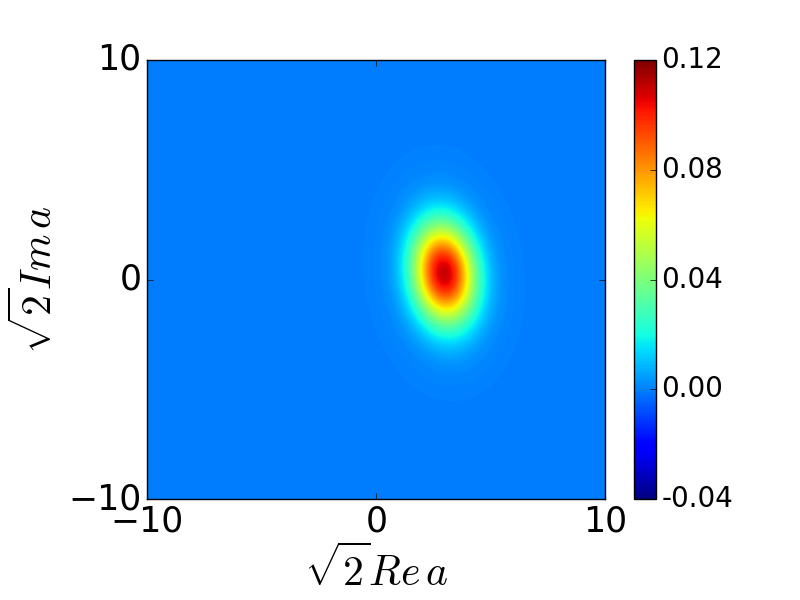}
             \includegraphics[width=0.32\textwidth]{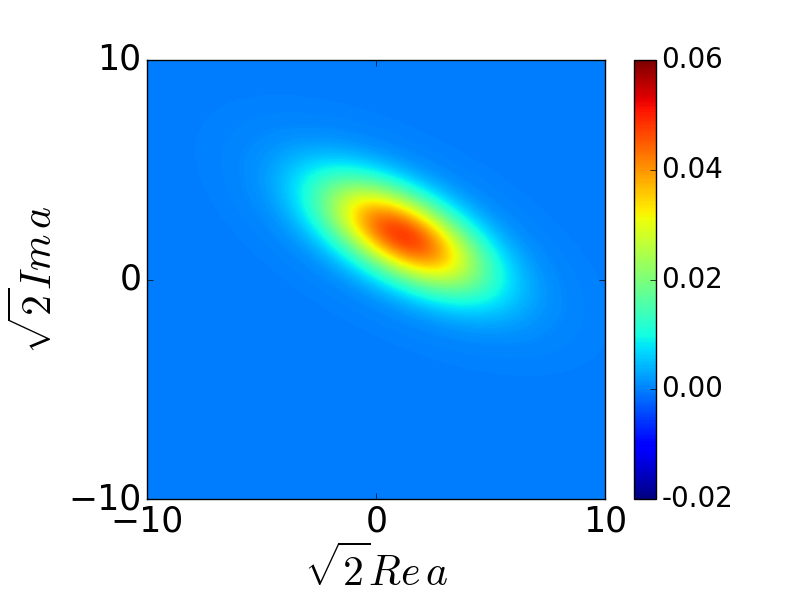}
              \includegraphics[width=0.32\textwidth]{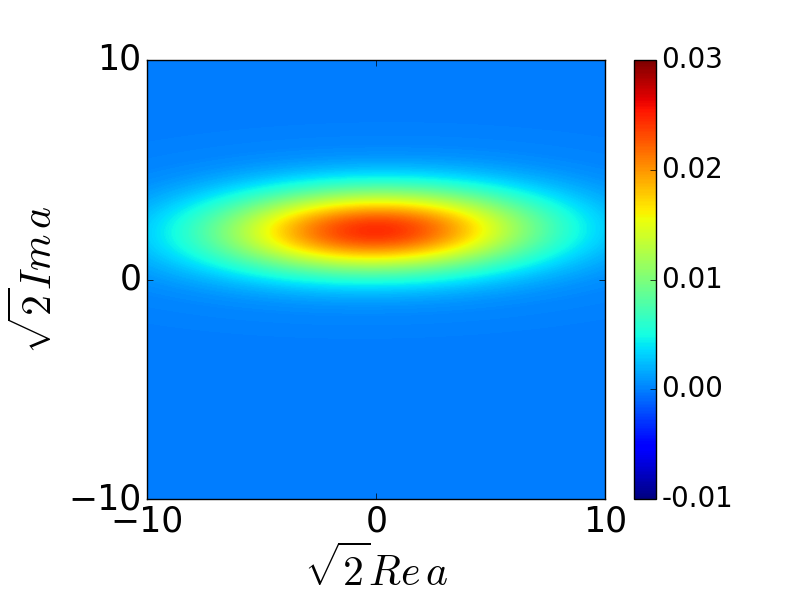}
        \caption{Comparison of the quantum (top row) and semiclassical (bottom row) dynamics of an initial Glauber coherent state with $\sqrt{2}\Re\,a=4$ and $\sqrt{2}\Im\,a=4$. The harmonic oscillator frequency $\omega=1$, while the linear damping rate, nonlinear damping rate and amplification rate are $\gamma_1=0.1$, $\gamma_2=0.01$ and $A=0.15$ respectively. Times $t= 13$, $50$, $150$ are shown left to right.} 
        \label{fig:wigner limit plot}
    \end{figure}
    
As discussed in \cite{Bowe16} the origin $|a|=0$ is a stable fixed point provided $A-\gamma_1<0$. However, when the value of $A$ exceeds $\gamma_1$ the system exhibits a Hopf bifurcation and the origin becomes unstable. In this case a stable limit cycle occurs and the long-term solution tends towards the curve $|a|^2 = (A-\gamma_1)/2\gamma_2$. A phase-space portrait for the case $A>\gamma_1$ is illustrated in Figure \ref{fig:limit pp} and the semiclassical and quantum dynamics are compared in Figure \ref{fig:wigner limit plot}. Due to the weak nonlinear damping a good correspondence is observed for short to medium times. Over longer times the centre of the semiclassical Gaussian becomes trapped on the limit cycle and the width grows. On the other hand, in accordance with the semiclassical flow (Figure \ref{fig:limit pp}), the quantum Wigner function begins to smear out over the limit cycle into a 'donut' shape.

\begin{figure}[htb]
      \centering
                 \includegraphics[width=0.49\textwidth]{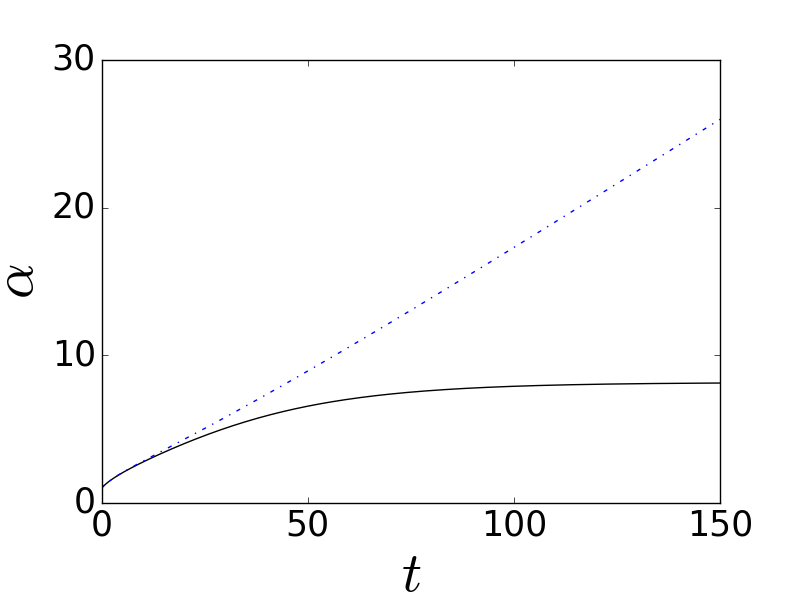}
 	\includegraphics[width=0.49\textwidth]{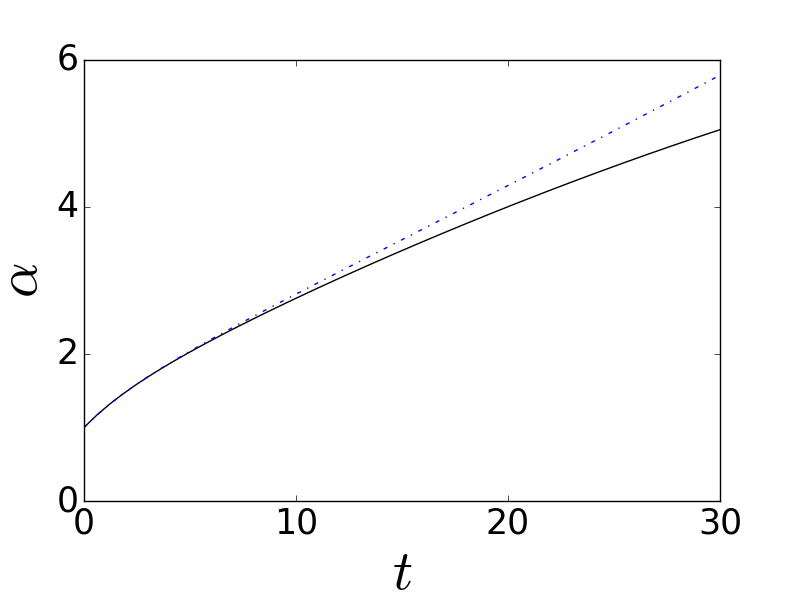}
\caption{The long-time evolution of the covariance element $\alpha = \la \hat{a}^\dag \hat{a} + \hat{a} \hat{a}^\dag \ra - 2 \la \hat{a}^\dag \ra \la \hat{a} \ra$ is plotted on the left. A close up of the dynamics at early times is shown on the right. The quantum dynamics (black) are compared to the semiclassical results (blue dashed), with the same initial conditions and parameters as Figure \ref{fig:wigner limit plot}.} 
\label{fig:covar dyn}
\end{figure}

In Figure \ref{fig:covar dyn} we additionally examine the time evolution of the covariance element $\alpha$, where $\alpha$ is defined in (\ref{eqn:alpha}). As expected, in the short time limit there is a good agreement with the quantum dynamics. Over a longer time the Gaussian continues to spread out, resulting in a growing value of $\alpha$. This is in contrast to the quantum value of $\alpha$, which eventually plateaus. This feature appears to stem from the quantum Wigner function smearing out over the limit cycle. In order to describe the spread of the Wigner function over a limit cycle in a semiclassical framework we would have to adapt the methods and results from \cite{SchuValTos12} to open systems. 

\hfill

As a second application, let us consider the description of cold atoms in optical lattices with particle losses \cite{Daley,Kordas}. In a full many-particle treatment for low energies the $M$-mode Bose-Hubbard Hamiltonian describes the unitary dynamics of ultracold atoms in an $M$-well optical trap
\begin{equation}
\label{eqn:Bose Hubbard}
\hat H = -J\sum_{j=1}^{M-1}(\hat{a}_j^\dagger \hat{a}_{j+1} + \hat{a}_{j+1}^\dagger \hat{a}_j) + \frac{U}{2}\sum_{j=1}^{M} \hat{a}_j^\dagger \hat{a}_j^\dagger \hat{a}_j \hat{a}_j.
\end{equation}
The $\hat{a}_j$ $(\hat{a}_j^\dagger)$ create (annihilate) particles in the ground state of the $i$\textsuperscript{th} trap, $J$ is the positive coupling constant and $U$ describes pairwise on-site interactions. Here we consider repulsive interactions with $U>0$. The two-body interaction is valid for cold dilute gases in which s-wave scattering dominates and higher order interactions can be ignored. 

The ground state of a Bose-Einstein condensate (BEC) in an optical lattice potential can be approximated as a product state of Glauber coherent states
\begin{equation}
\label{eqn:BEC state}
|\psi\ra = \bigotimes_i |\alpha_i\ra, 
\end{equation}
which we take as our initial state. The on-site Glauber coherent state $|\alpha_i\ra$ is defined as an eigenstate of the annihilation operator $\hat a_i$, where the complex eigenvalue $\alpha_i$ describes the amplitude and phase of the coherent matter wave-field. This corresponds to a Poissonian atom number distribution on each site with average  $\bar n_i = |\alpha_i|^2$. Thus, the atom number at each site is uncertain. 

The celebrated mean-field description of BECs in optical lattices can be derived using a Gaussian approximation. In fact, this goes beyond the mean-field description as it also provides an approximation for correlation functions. While the mean-field dynamics for cold atoms in optical lattices for specific Lindblad operators have previously been derived in the literature by \textit{neglecting quantum fluctuations} in the equations of motion for particular observables, the Gaussian approximation introduced here yields a canonical form of the mean-field dynamics for arbitrary Lindblad operators. We shall now provide these equations of motion for the general case, before briefly discussing the resulting dynamics for an example system. 

\begin{figure}[h]
      \centering
        \includegraphics[width=0.49\textwidth]{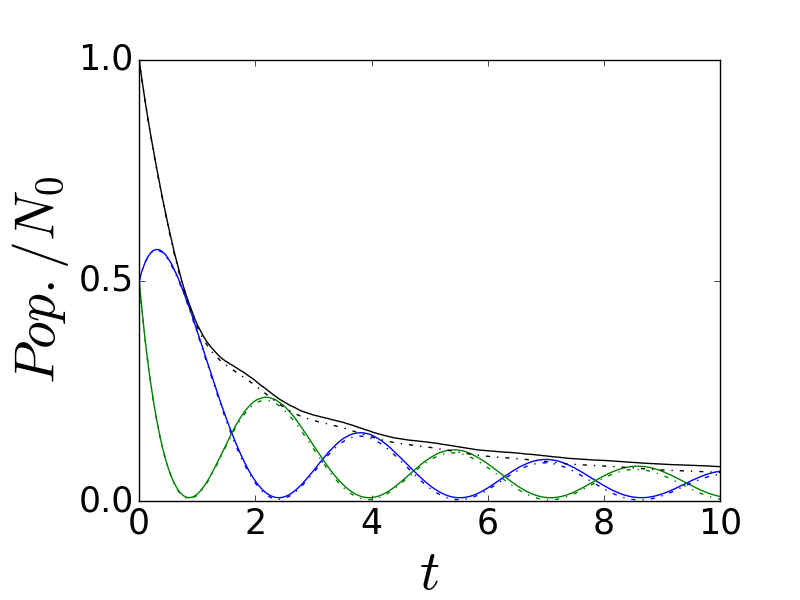}
         \includegraphics[width=0.49\textwidth]{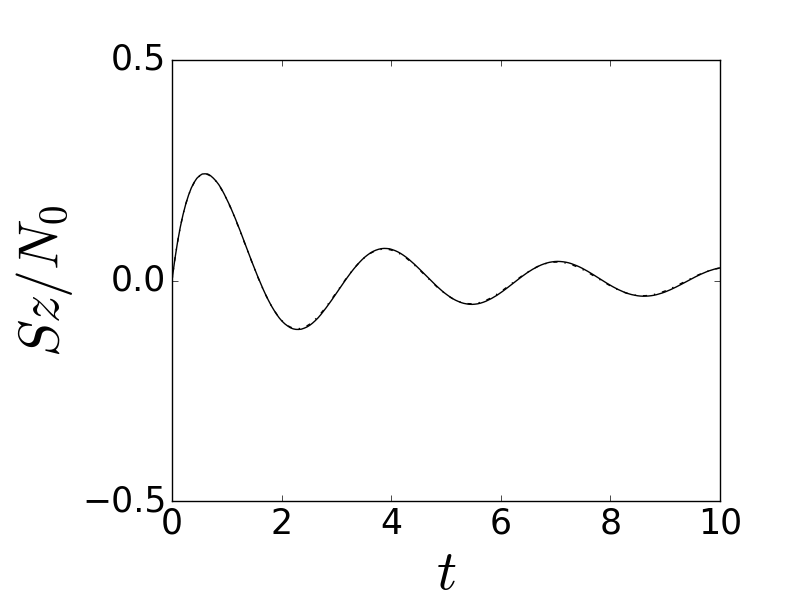}
\caption{The left hand frame compares the quantum (solid lines) and semiclassical (dashed lines) dynamics of the total particle number (black), as well as the populations of site 1 (blue) and site 2 (green). The quantum (solid line) and semiclassical (dashed line) population imbalance is plotted in the right frame. In both cases $J=1$, the initial particle number $N_0=20$, $UN_0 = 1$, $\gamma = 0.05$ and the particles are initially distributed evenly across both sites, with phases $\phi_1 = \pi/2$ and $\phi_2 = 0$. The quantum results were obtained using the quantum jump method.}
\label{fig:numbers two body losses} 
\end{figure}

The Weyl symbol of the Bose-Hubbard Hamiltonian (\ref{eqn:Bose Hubbard}) is given by
\begin{equation}
\label{eqn:BH Weyl}
H = -J\sum_{j=1}^{M-1} \left( \bar{a}_j a_{j+1} + \bar{a}_{j+1} a_j \right) + \frac{U}{2} \sum_{j=1}^{M} \left( |a_j|^4 - 2|a_j|^2 + \frac{1}{2} \right).
\end{equation}
Inserting the Hamiltonian above into (\ref{eqn:Xc eom}) yields the Gross-Pitaevskii equation describing the Hamiltonian dynamics together with additional terms describing the effects of the Lindblad operators
\begin{equation}
\label{eqn:GP open}
\dot{a}_j = iJ(a_{j+1} + a_{j-1}) - iU(|a_j|^2 - 1) a_j + \frac{1}{2}\sum_k \left(\bar{L}_k \frac{\pa L_k}{\pa \bar{a}_j} - L_k \frac{\pa \bar{L}_k}{\pa \bar{a}_j} \right).
\end{equation}
Thus, the usual mean-field result is recovered, together with the effect of any Lindblad operator that is a function of $\hat{a}$ and $\hat{a}^\dagger$. Furthermore, the equation of motion for $G_c$ provides short-time approximations for quantities depending on the covariance matrix elements. For instance, the phase-coherence of the state is characterised by the first-order correlation function
\begin{equation}
\label{eqn:first cor}
g_{ij}^{(1)} = \frac{|\la \hat{a}_i^\dag \hat{a}_j \ra|}{\sqrt{\la \hat{a}_i^\dag \hat{a}_i \ra \la \hat{a}_j^\dag \hat{a}_j \ra}}.
\end{equation} 
Let us consider a two-mode system with losses due to two-body inelastic scattering, which can be modelled by applying the nonlinear Lindblad operators $\hat{L}_j = \sqrt{\gamma}\hat{a}_j^2$ to each lattice site \cite{Daley,Kordas}. In this case (\ref{eqn:GP open}) leads to the equations of motion
\begin{align}
\dot{a}_1 &= iJa_2 - iU\left(|a_1|^2-1\right)a_1 - \gamma |a_1|^2 a_1, \label{eqn:a1 eom} \\
\dot{a}_2 &= iJa_1 - iU\left(|a_2|^2-1\right)a_2 - \gamma |a_2|^2 a_2. \label{eqn:a2 eom}
\end{align}
The solution of these equations provides a good first order approximation to the total population $N(t) = |a_1(t)|^2+|a_2(t)|^2$ and the population imbalance $2S_z(t) = |a_1(t)|^2-|a_2(t)|^2$, as illustrated in Figure \ref{fig:numbers two body losses} for an example with relatively small loss rate.

\begin{figure}[h]
      \centering
        \includegraphics[width=0.49\textwidth]{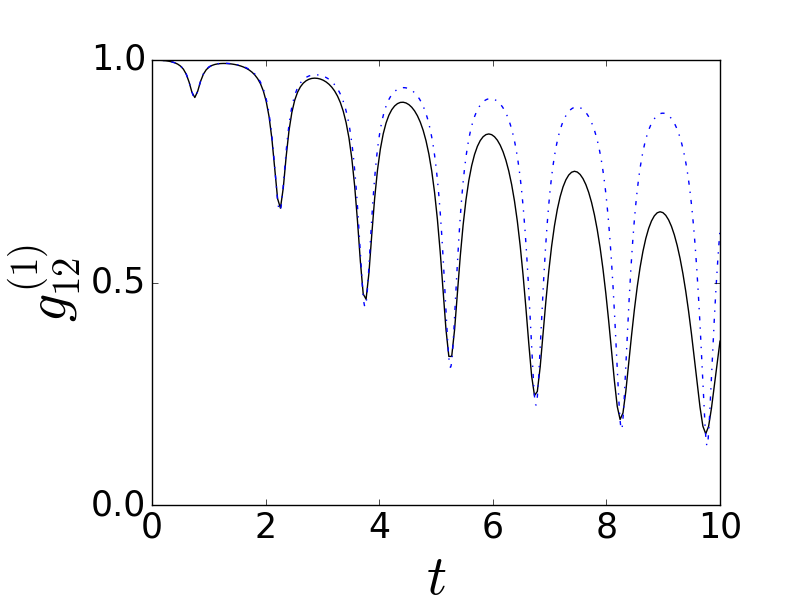}
         \includegraphics[width=0.49\textwidth]{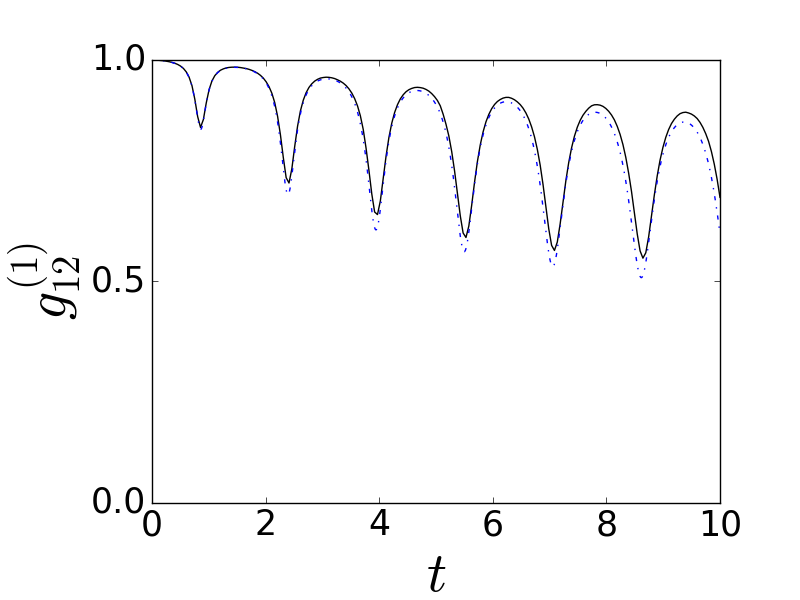}
\caption{The quantum (black) and semiclassical (blue dashed) dynamics of the phase-coherence $g_{12}^{(1)}$ between the two traps, for a closed (left) and open system with $\gamma = 0.05$ (right). In both cases $J=1$, the initial particle number $N_0=20$, $UN_0 = 1$ and the particles are initially distributed evenly between the two traps, with phases $\phi_1 = \pi/2$ and $\phi_2 = 0$ respectively. The quantum results were obtained using the quantum jump method.}
\label{fig:phase two body losses} 
\end{figure}

In the right panel of Figure \ref{fig:phase two body losses} we compare the quantum and semiclassical dynamics of the phase-coherence for the same system as in Figure \ref{fig:numbers two body losses}. In the left panel we show the same comparison for the closed system for reference. As expected, a loss of phase-coherence is observed over time, indicating destruction of the condensate. The inclusion of losses leads to a better correspondence between the semiclassical and quantum dynamics for short times. This can be understood to some extent by the fact that since the Lindblad operators are holomorphic functions of $\hat a_j$  the dissipative parts of (\ref{eqn:a1 eom}) and (\ref{eqn:a2 eom}) may be written as a gradient flow. In the semiclassical limit the effect of the Lindblad operators is thus to localise the Wigner function. 

\section{Lindblad Dynamics as Schr\"{o}dinger Dynamics with a Non-Hermitian Hamiltonian}\label{section:non herm}
In this section we approach the Lindblad dynamics from a different perspective, specifically, by interpreting the Lindblad equation as a Schr\"{o}dinger equation with a non-Hermitian Hamiltonian. The notion of a phase-space Schr\"{o}dinger equation and phase-space wave functions is not new. Notably, for closed systems, Koda recently formulated several initial value semiclassical propagators for the Wigner function starting from an interpretation of the Moyal equation as a Schr\"{o}dinger equation \cite{Koda15}. In the case of Lindblad dynamics, the new ingredient is the non-Hermiticity of the resulting Hamiltonian operator. Quantum dynamics generated by non-Hermitian Hamiltonians has in recent years attracted considerable attention in its own right (see, e.g., \cite{Mois11book,Bend07}, and references therein). In \cite{Grae11,Grae12} two of the authors have developed the semiclassical limit of Gaussian wave packet propagation for non-Hermitian Hamiltonians. Reinterpreting the Wigner-evolution for Lindblad systems as a non-Hermitian Schr\"odinger equation, we can directly apply these results to obtain dynamical equations that can describe the semiclassical propagation of complex Gaussian Wigner functions, as they appear for example in the superpositions of Gaussian wave packets. 

Let us for example consider an initial state $\phi(q)=\sum_j c_j \phi_j^A (q)$ that is a superposition of Gaussian wave packets
\begin{equation}
\label{eqn:gauss packet}
\phi_j^A (q) = \frac{(\det\Im A)^{1/4}}{(\pi\hbar)^{n/4}} \ue^{\frac{i}{\hbar} [\frac{1}{2}(q-q_j)\cdot A(q-q_j) + p_j\cdot(q-q_j)]},
\end{equation}
 where $q, q_j,p_j \in \R^n$ and $A$ is an $n\times n$ complex symmetric matrix with $\Im A >0$. The Wigner function of this state is
 \begin{equation}
 \label{eqn:wigner sup}
 W(x) = \sum_{i,j} c_i^*c_j \psi_{ij}(x),
 \end{equation}
where $x = (q,p)$ and
\begin{equation}
\label{eqn:wig component}
\psi_{ij}(x) = \frac{1}{\left(\pi\hbar\right)^n} \ue^{\frac{i}{\hbar}\left[(x-X_{ij})\cdot iG (x-X_{ij}) + Y_{ij}\cdot (x-X_{ij}) + \alpha_{ij}\right]}
\end{equation}
is a complex Gaussian centred at $X_{ij} = \frac{1}{2}\left(q_i+q_j,p_i+p_j\right)$ with 'momentum' $Y_{ij} = \left(p_j-p_i,q_i-q_j \right)$ and a complex phase $\alpha_{ij} =  \frac{1}{2} (p_i + p_j)\cdot (q_i - q_j)$. The matrix $G$ is related to the width $A$ of the Gaussians in the superposition by
 \begin{equation}
 \label{eqn:G def B}
 G = \begin{pmatrix} \Im A +\Re A [\Im A]^{-1}\Re A  & -\Re A[\Im A]^{-1}\\  -[\Im A]^{-1}\Re A & [\Im A]^{-1} \end{pmatrix}.
 \end{equation}
As the phase-space Lindblad equation (\ref{eqn:wigner moyal eom}) is linear in the Wigner function $W$, the time evolution of  (\ref{eqn:wigner sup}) can be obtained by evolving each complex Gaussian $\psi_{ij}$ individually and summing the results. 

With this picture in mind our starting point is once again the Lindblad equation on phase space
\begin{equation}
\label{eqn:wigner moyal eom 2}
i\hbar \frac{\pa \psi}{\pa t} = H\star \psi-\psi\star H+i\sum_k L_k\star \psi\star\bar{L}_k - \frac{1}{2} \bar{L}_k\star L_k\star \psi - \frac{1}{2}\psi\star \bar{L}_k\star L_k.
\end{equation}
However, we have now switched notation from $W$ to $\psi$ to indicate that we could be dealing with a complex component of the Wigner function, such as the $\psi_{ij}$ in the discussion above. As this equation is linear in $\psi$ we can in fact view (\ref{eqn:wigner moyal eom 2}) as a Schr\"{o}dinger equation with a non-Hermitian Hamiltonian in a larger Hilbert space.

To this end we will use the fact that the star product can be written as the action of an operator on phase-space functions, see e.g. \cite{Koda15}. More precisely, we have 
\begin{equation}
\label{eqn:star to op}
A \star \psi = \hat A^{(-)} \psi \quad \textnormal{and} \quad \psi \star A = \hat A^{(+)} \psi,
\end{equation}
where the operators
\begin{equation}
\label{eqn:double op}
\hat A^{(\pm)} = A(\hat x\pm \frac{1}{2} \Omega \hat y)
\end{equation}
act on the phase-space function $\psi(x)$. Furthermore,
\begin{align}
\label{eqn:double xy}
\hat x &= (q,p),\\
\hat y &= \left(-i\hbar\nabla_q,-i\hbar\nabla_p\right),
\end{align}
are a pair of Hermitian operators that satisfy the canonical commutation relations
\begin{equation}
\label{eqn:double canonical}
[\hat x_i,\hat y_j] = i\hbar\delta_{ij}
\end{equation}
and can thus be treated like position and momentum operators in a space of doubled dimension. Making use of (\ref{eqn:star to op}), the phase-space Lindblad equation (\ref{eqn:wigner moyal eom 2}) can be written as a Schr\"{o}dinger equation
\begin{equation}
\label{eqn:non herm eom}
i\hbar \frac{\pa W(x,t)}{\pa t} = \hat K (\hat x, \hat y) W(x,t),
\end{equation} 
with non-Hermitian Hamiltonian 
\begin{equation}
\label{eqn:non herm ham}
\hat K(\hat x,\hat y) = \hat H^{(-)} - \hat H^{(+)} + i\sum_k \hat L^{(-)}_k \hat{\bar{L}}^{(+)}_k - \frac{1}{2}\widehat{(\bar{L}_k \star L_k)}^{(-)} - \frac{1}{2}\widehat{(\bar{L}_k \star L_k)}^{(+)}.
\end{equation}
Note that this is a phase-space analogue of a matrix-vector representation of the Lindblad equation (see e.g. in \cite{NEFT1,NEFT2}).

Finally, through the Wigner-Weyl transformation the non-Hermitian Hamiltonian $\hat K$ can be mapped onto the double phase-space function 
\begin{align}
\label{eqn:non herm symb}
K(x,y) &= H\big(x-\frac{1}{2}\Omega y \big)-H\big(x+\frac{1}{2}\Omega y\big) + i \sum_{k} L_k\big(x-\frac{1}{2}\Omega y\big) \star_2 \bar{L}_k\big(x+\frac{1}{2}\Omega y\big) \nonumber \\ &-\frac{1}{2}(\bar{L}_k \star L_k)\big(x-\frac{1}{2}\Omega y\big) - \frac{1}{2}(\bar{L}_k\star L_k)\big(x+\frac{1}{2}\Omega y\big),
\end{align}
where $\star_2$ denotes the Moyal product on the double phase-space. If we assume that the Hamiltonian and Lindblad operators are quantisations of $\hbar$-independent functions  $H$ and $L_k$,   then using \eqref{eqn:Moyal product1} we find that  $K$ has a semiclassical expansion  $K=K^{(0)}+\hbar K^{(1)}+\dots$, where the first two terms are given by 
\begin{align}
K^{(0)} & =H^{(+)}-H^{(-)}+\sum_k\Im \big(\bar L_k^{(-)} L_k^{(+)} \big) -\frac{\ui}{2}\sum_k\big|L_k^{(+)}- L_k^{(-)}\big|^2\,\, ,\label{eq:im K} \\
K^{(1)} & =\frac{1}{2}\sum_k \{\bar L_k,L_k\}^{(+)}+\{\bar L_k,L_k\}^{(-)}\,\, .\label{eq:K1} 
\end{align}
In the expression for $K^{(0)}$ we have separated real and imaginary parts, whereas $K^{(1)}$ is imaginary.\footnote{We can allow more general functions which have asymptotic expansions in powers of $\hbar$, e.g., $H\sim H^{(0)}+\hbar H^{(0)}1+\cdots $ and similarily for $L_k$. Then $K^{(0)}$ is determined by $H^{(0)}$ and $L_k^{(0)}$ only and $K^{(1)}$ contains further terms depending on $H^{(1)}$ and  $L_k^{(1)}, L_k^{(0)}$.}

We are now in the position to study the time evolution of initial wave packets of the form 
\begin{equation}
\label{eqn:wigner wave packet}
\psi(x) = \frac{(\det \Im B)^{1/4}}{(\pi\hbar)^{n/2}}\ue^{\frac{i}{\hbar}\left[(x-X)\cdot B(x-X)/2 + Y\cdot(x-X) +\alpha \right]},
\end{equation}
where $X,Y \in \R^{2n}$, $\alpha \in \C$ is a phase factor and $B$ is a complex symmetric matrix with $\Im B > 0$. As shown above, such states appear as components of the Wigner function of superpositions of Gaussian states (\ref{eqn:wigner sup}). 

Direct application of the semiclassical equations derived in \cite{Grae11,Grae12} yields the following equations of motion 
\begin{align}
\dot Z &= \Omega_2 \nabla \Re K^{(0)} + \mathcal{G}^{-1} \nabla \Im K^{(0)}, \label{eqn:Z dbl eom}\\
\dot B &=  - BK_{yy}^{(0)}B - BK_{y x} ^{(0)}- K_{x y}^{(0)}B  - K_{xx}^{(0)},\label{eqn:B eom}\\
\dot \alpha &= \frac{i\hbar}{4}\Tr(\dot B B^{-1}) + Y\cdot \dot X - K^{(0)}(X,Y)-\hbar K^{(1)}(X,Y) \nonumber\\
&\phantom{==} +\frac{i\hbar}{2}\Tr(K^{(0)}_{xy} + K^{(0)}_{yy}B) \label{eqn:phase db},
\end{align}
where $Z = (X,Y)$, $\nabla \defeq \left(\pa_x,\pa_y\right)$ is the double phase-space gradient, and $\Omega_2$ is the double phase-space symplectic form. We have also defined
\begin{equation}
K_{xy} = (\pa_{x_i} \pa_{y_j}K),
\end{equation}
which satisfies $K_{y x} = \left(K_{xy}\right)^\Tp$, and $\mathcal{G}$ is related to $B$ via 
\begin{equation}
\label{eqn:g def B}
\mathcal{G} = \begin{pmatrix} \Im B+\Re B [\Im B]^{-1}\Re B  & -\Re B[\Im B]^{-1}\\  -[\Im B]^{-1}\Re B & [\Im B]^{-1} \end{pmatrix}.
\end{equation}

In order to obtain some insight into the properties of these equations of motion we use the specific form of the real and imaginary parts of $K$ in \eqref{eq:im K}. 
We see that the imaginary part is an even function of $y$, $\Im  K^{(0)} (x,-y)=\Im  K^{(0)} (x,y)$, with $\Im  K^{(0)} (x,0)=0$, and non-positive.  
The real part is an odd function of $y$, $\Re  K^{(0)} (x,-y)=-\Re^{(0)}  K (x,y)$. Using these properties and direct computations we find that
\begin{align}
\nabla \Re K^{(0)}(x,0) &=\bigg(0, \Omega\nabla_xH(x)+\Omega \sum_k \Im\big(L_k(x)\nabla \bar L_k(x)\big)\bigg)\,\, ,\\
\nabla \Im K^{(0)} (x,0) & =\big(0,0\big).
\end{align}
By inserting this result into \eqref{eqn:Z dbl eom} we observe that if the initial value of $Y$ is $0$ then Y stays $0$ for all times, and $X$ satisfies the same equation of motion we found in \eqref{eqn:X eom}. 
If we insert  $B= 2iG$ into \eqref{eqn:B eom}, and separate the real and imaginary parts of the Hessian matrix of $K^{(0)} $ at $y=0$ in the same way, we find similarly that 
$G$ satisfies \eqref{eqn:G eom}. Hence our two different approaches are consistent. 

The next natural question to ask is what happens when $Y\neq 0$. In this case we have a highly oscillatory initial Wigner function, which corresponds to a very non-classical state.  As the imaginary part $\Im K^{(0)} (x,y)$ in \eqref{eq:im K} has a maximum at $y=0$, we see that the gradient part in \eqref{eqn:Z dbl eom} wants to push 
$Y$ to $0$, and hence reduce the frequency of the oscillations. In addition, \eqref{eqn:phase db} generates an exponentially damping factor if $\Im K^{(0)} (x,y) <0$, and both these effects are directly induced by the 
Lindblad operators. Thus, the general structure  of the equations of motion \eqref{eqn:Z dbl eom} and \eqref{eqn:phase db}, together with \eqref{eq:im K}, allow us to conclude that oscillatory initial conditions will 
be smoothed and suppressed exponentially fast if they couple to the Lindblad terms via \eqref{eq:im K}. This is a manifestation of decoherence \cite{Schl10}, which is illustrated in an example at the end of this section.


Before providing an example we compare our results to the work of Brodier and Ozorio de Almeida \cite{Brod10b}, who have also studied the Lindblad evolution of Gaussian states but for the special case of linear Lindblad operators of the form $\hat{L}_k =l_k\cdot \hat x$, where $l\in \C^{2n}$. We shall see that additional terms appear if one goes beyond the linear case. Brodier and Ozorio de Almeida consider the chord function $\chi_{\hat{\rho}} (t,y):=\Tr[\hat{\rho}\ue^{-\frac{\ui}{\hbar} y\cdot \hat{x}}]$ instead of the Wigner function. As the chord function is the Fourier transform of the Wigner function, the chord function of \eqref{eqn:wigner wave packet} is given by
\begin{equation}
\chi(y)=N\ue^{\frac{\ui}{\hbar}[\frac{1}{2}(y-Y)\cdot (\mathbb{N}+\ui\mathbb{M})(y-Y)+X\cdot y]}.
\end{equation}
Here $\mathbb{N},\mathbb{M}$ are real symmetric with $-B^{-1}=\mathbb{N}+\ui\mathbb{M}$ and the normalisation $N$ can be determined from $\chi(0)=\Tr \hat\rho$. Note that the additional phase factors in the Gaussian have been included in $N$ for convenience. We now want to show that the equations of motions for $X,Y$,$\mathbb{N}$ and $\mathbb{M}$ match those in \cite{Brod10b}. To this end we use that 
\begin{equation}
\mathcal{G}^{-1}=\begin{pmatrix} \mathbb{M}+\mathbb{N} \mathbb{M}^{-1}\mathbb{N} & -\mathbb{N}\mathbb{M}^{-1} \\ -\mathbb{M}^{-1} \mathbb{N} & \mathbb{M}^{-1} \end{pmatrix},
\end{equation}
which one may find by direct computation, or more easily by applying Proposition 3.2 in \cite{Grae12} with $S=\Omega^{-1}$. For linear Lindblad operators $\Im K (x,y)=-\frac{1}{2} y\cdot \mathbb{D} y$, with $\mathbb{D} =\sum_k \Re(\bar{l}_k l_k^T) $, and the equation of motion for $Z$ \eqref{eqn:Z dbl eom} may be rewritten as 
\begin{equation}\label{eq:BrodXY}
\dot X=\nabla_y\Re K +\mathbb{N}\mathbb{M}^{-1}\mathbb{D} Y\,\, \quad  \dot Y=-\nabla_x\Re K  -\mathbb{M}^{-1}\mathbb{D} Y.
\end{equation}
The equations of motion for $\mathbb{M}$ and $\mathbb{N}$ are obtained by using $\mathbb{N}+\ui\mathbb{M}=-B^{-1}$, which gives $\dot{\mathbb{N}}+\ui\dot{\mathbb{M}}=B^{-1}\dot{B} B^{-1}$. After inserting \eqref{eqn:B eom}, and separating the real and imaginary parts, we find 
\begin{align}
\dot{\mathbb{M}}&=\mathbb{D}+\Re K_{yx} \mathbb{M}+\mathbb{M} \Re K_{xy}-\mathbb{M}\Re K_{xx}\mathbb{N}-\mathbb{N}\Re K_{xx}\mathbb{M}\label{eq:BrodM}\\
\dot{\mathbb{N}}&=-\Re K_{yy}+\Re K_{yx} \mathbb{N}+\mathbb{N} \Re K_{xy}+\mathbb{M}\Re K_{xx}\mathbb{M}-\mathbb{N}\Re K_{xx}\mathbb{N}\,\, .\label{eq:BrodN}
\end{align}
The equations \eqref{eq:BrodXY}, \eqref{eq:BrodM} and \eqref{eq:BrodN} coincide with the ones derived in \cite{Brod10b}. The main generalisation in our approach is that it can account for more general Lindblad operators, which need no longer be linear 
functions of $\hat x$. For general Lindblad operators $\Im K^{(0)}$ will depend on both $y$ \textit{and} $x$, leading to additional terms in the equations of motion for $X,Y, \mathbb{N}$ and $\mathbb{M}$ compared to those in \cite{Brod10b}. This can be observed in \eqref{eqn:G eom}, where the equation for $G=\mathbb{M}^{-1}/2$ contains terms involving second derivatives of the Lindblad operators. 
\\

We finally show how the doubled phase-space approach allows for a treatment of interference terms in the Wigner function by obtaining the semiclassical dynamics of an initial Schr\"{o}dinger cat state in a damped anharmonic oscillator. Working in dimensionless units with $\hbar = 1=m=\omega$, the Hamiltonian is
\begin{equation}
\label{eqn:H anharm}
\hat H = \frac{1}{2}(\hat q^2 + \hat p^2) + \frac{\beta}{4}\hat q^4,
\end{equation}
where $\beta$ controls the degree of anharmonicity, and the damping is modelled with the Lindblad operator 
\begin{equation}
\hat L = \sqrt{\frac{\gamma}{2}}(\hat q + i \hat p),
\end{equation}
where $\gamma$ determines the damping rate. When $\beta=0$ the semiclassical dynamics are exact. The corresponding double phase-space symbol $K$ is found from (\ref{eqn:non herm symb}) to be
\begin{equation}
K^{(0)}(x,y) = (\Omega x)\cdot y - \frac{\beta}{4}(x_q y_p^3 + 4 x_q^3 y_p) - \frac{\gamma}{2}x\cdot y - \frac{i\gamma}{4}y\cdot y,
\end{equation}
and 
\begin{equation}
K^{(1)}(x,y) =  \frac{i\gamma}{2}.
\end{equation}

\begin{figure}[htb]
      \centering
                 \includegraphics[width=0.25\textwidth]{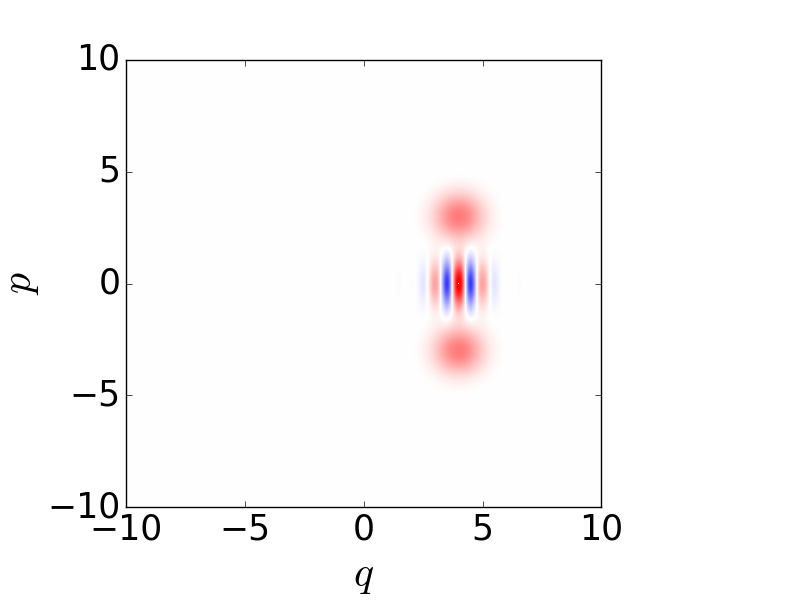}   
                 \hspace{-0.25cm} 
                  \includegraphics[width=0.25\textwidth]{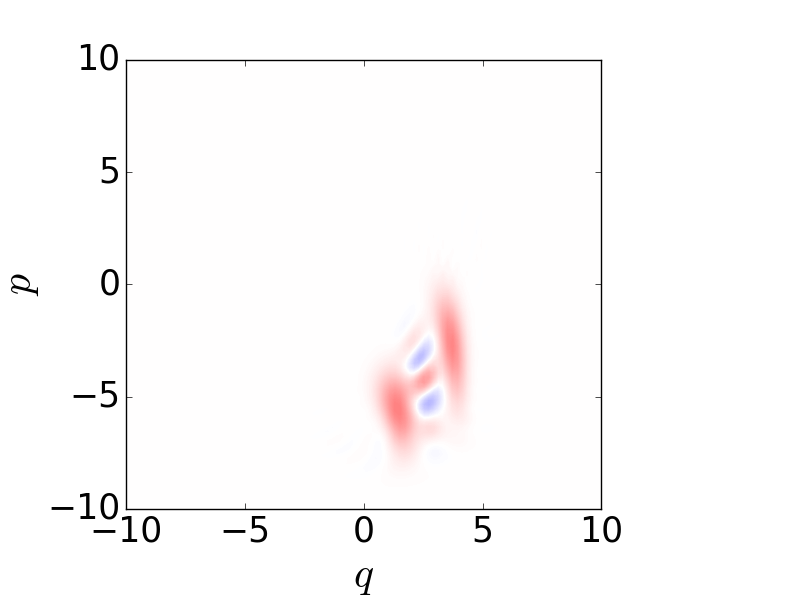}
                  \hspace{-0.25cm} 
                   \includegraphics[width=0.25\textwidth]{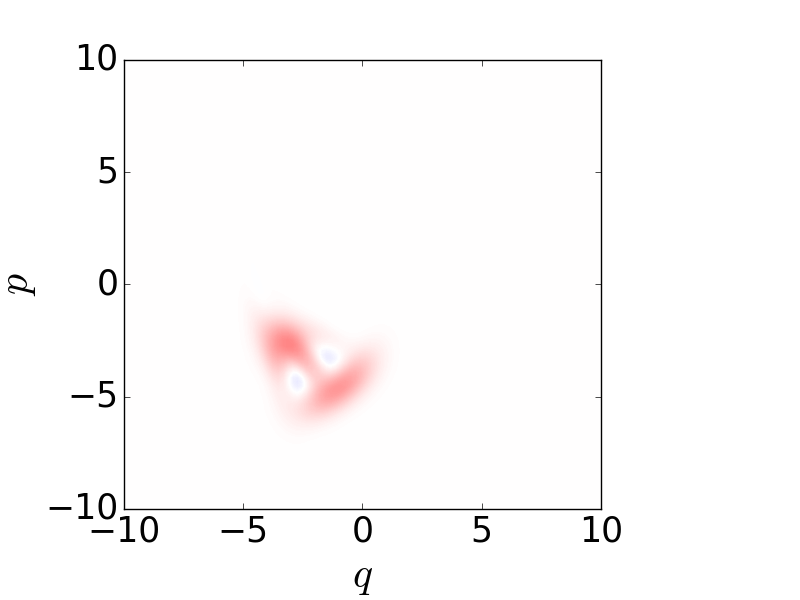}
                      \hspace{-0.25cm} 
                              \includegraphics[width=0.25\textwidth]{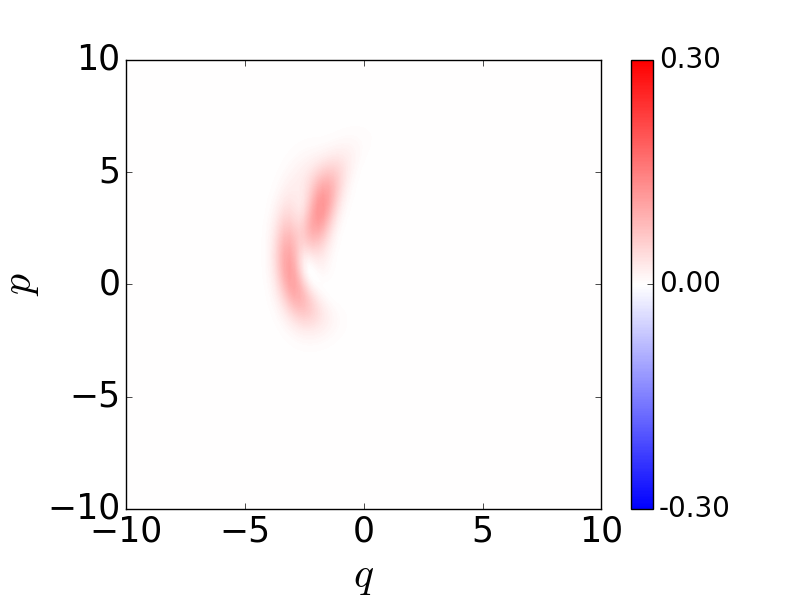}  
 	\includegraphics[width=0.25\textwidth]{cat_init.png}
	   \hspace{-0.25cm} 
                \includegraphics[width=0.25\textwidth]{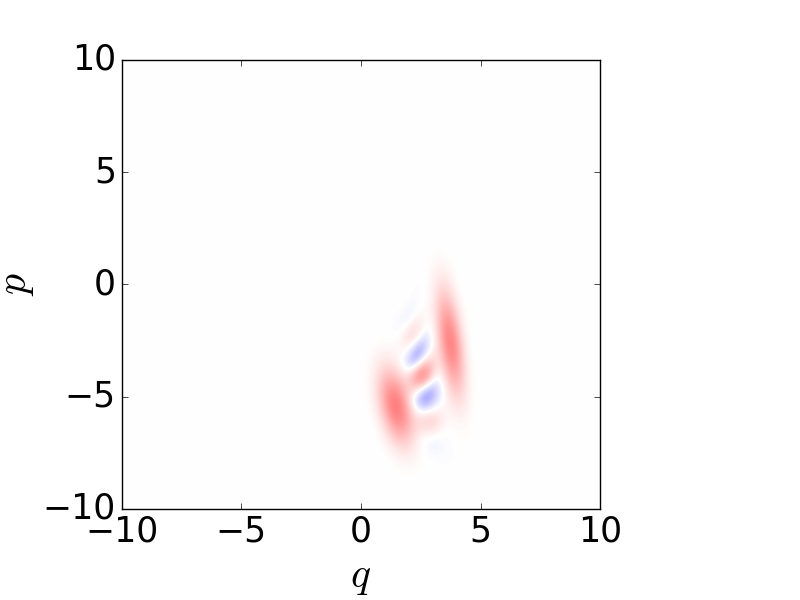}
                   \hspace{-0.25cm} 
           \includegraphics[width=0.25\textwidth]{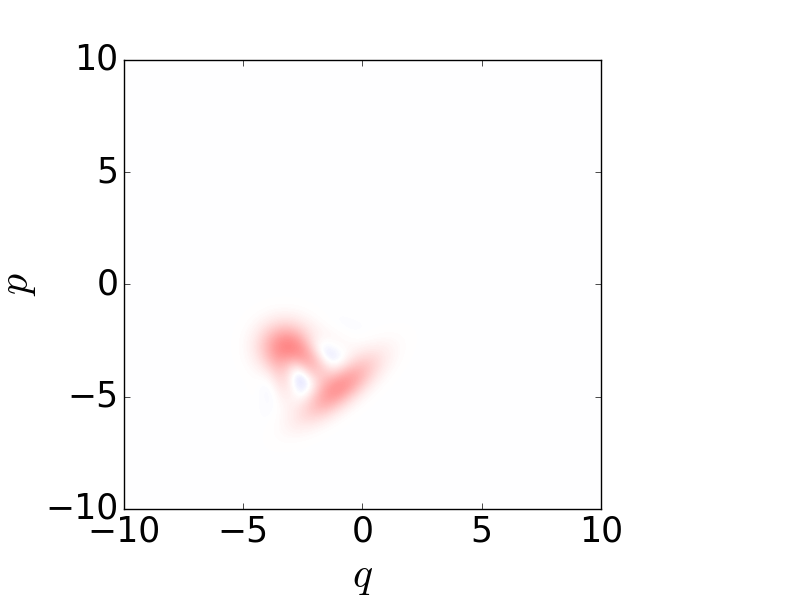}
              \hspace{-0.25cm} 
            \includegraphics[width=0.25\textwidth]{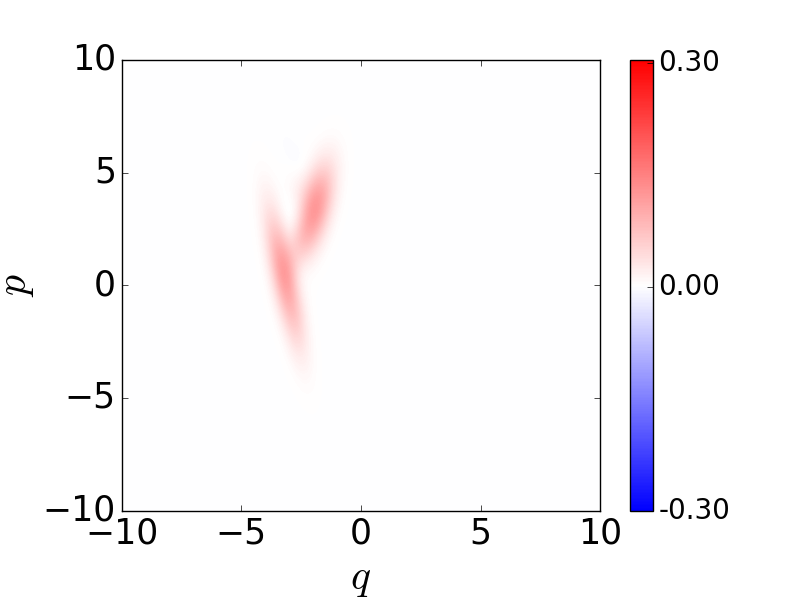}
\caption{The quantum (top) and semiclassical (bottom) dynamics of an initial cat state in an anharmonic potential with $\beta=0.1$ and damping at a rate $\gamma=0.3$. Times $t=0,\, 0.5,\ 1.5,\, 2.5$ are shown from left to right.} 
\label{fig:cat plots}
\end{figure}

We consider an initial state that is a cat state, comprised of two Gaussians with $A=1$ and both centred at $q = 4$ with momenta $p = \pm 3$. The Wigner function of this state (depicted on the left of Figure \ref{fig:cat plots}) is composed of two Gaussians centred at $(q = 4, p = \pm 3)$ in phase space, and an interference pattern centred at the midpoint of the two Gaussians with a Gaussian envelope. The semiclassical dynamics are obtained by evolving each complex Gaussian component of the initial Wigner function, summing up the results, and applying the Wigner function normalisation condition $\int dx W(x) = 1$. The resulting semiclassical dynamics are compared with the quantum dynamics in Figure \ref{fig:cat plots}. The Gaussian approximation clearly reproduces the essential features of the oscillation, damping and decoherence. 

In Figure {\ref{fig:cat expec} we further depict a comparison between the semiclassical position and momentum expectation values and the quantum results for a longer timescale. We find that there is good agreement at short times, while at longer times the results deviate as expected, while still capturing the qualitative features of the dynamics.
 
\begin{figure}[h]
      \centering
                 \includegraphics[width=0.49\textwidth]{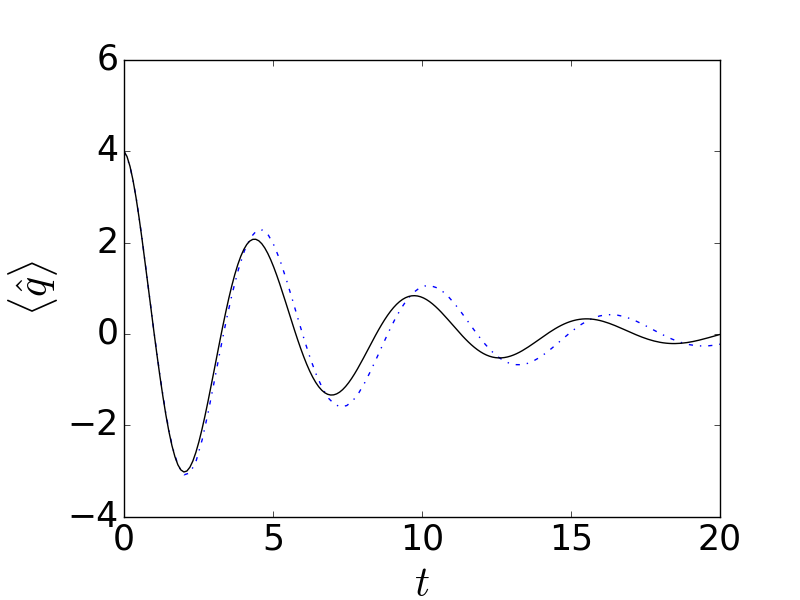}
 	\includegraphics[width=0.49\textwidth]{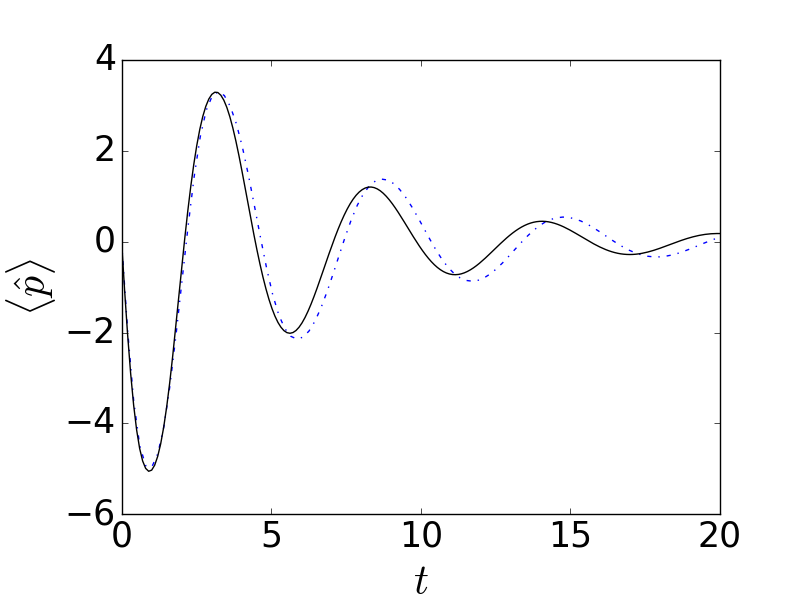}
\caption{Time evolution of the position (left) and momentum (right) expectation values of the cat state above. The quantum dynamics (black) are compared to the semiclassical results (blue dashed). Anharmonic parameter $\beta=0.1$ and damping rate $\gamma=0.3$.} 
\label{fig:cat expec}
\end{figure}

\section{Summary}

We have investigated the dynamics of Gaussian states in open quantum systems described by Lindblad equations in the semiclassical limit. This yields a new type of semiclassical phase-space dynamics incorporating the effects of damping and decoherence in the dynamics of the phase-space coordinates as well as a semiclassical approximation for the quantum covariances. This dynamics has an interesting geometric structure, which we have explored in a number of example systems. We have also transformed the dynamics to complex phase-space variables, as they appear naturally in many models in quantum optics. 

What makes the dynamics of Gaussian wave packets particularly appealing for closed quantum systems is the fact that an arbitrary initial wave function can be expanded into Gaussian states, each of which can be propagated independently. This expansion can also be repeated at intermediate time-steps, allowing for numerical quantum dynamics that can in principle be arbitrarily accurate \cite{Hube88b,Hube89}. Due to the fact that Lindblad dynamics generates mixed states, this approach cannot be directly generalised to the case of open systems considered here. However, we have demonstrated how the interpretation of the Wigner dynamics in phase space as a Schr\"odinger dynamics with non-Hermitian Hamiltonian can be used to circumvent this issue. We have demonstrated that this allows for the semiclassical propagation of interference terms by considering a cat state in a damped anharmonic oscillator. 

\section*{Acknowledgements}
E.M.G. acknowledges support from the Royal Society (Grant. No. UF130339) and from the European Research Council (ERC) under the European Union?s Horizon 2020 research and innovation programme (grant agreement No 758453). B.L. acknowledges support from the Engineering and Physical Sciences Research Council via the Doctoral Training Partnership (Grant No. EP/M507878/1).


\vspace{1cm}
\section*{References}
\begin{thebibliography}{10}

\bibitem{Schr26}
E.~Schr\"odinger,  {\it Der stetige \"Ubergang von der Mikro- zur Makrophysik},
   Naturwissenschaften  {\bf 14}  (1926)   664

\bibitem{Dira58}
P.~A.~M. Dirac,  {\em The Principles of Quantum Mechanics},   Oxford University
  Press, Oxford, 1958

\bibitem{Hell75}
E.~J. Heller,  {\it Time-dependent approach to semiclassical dynamics},  J.
  Chem. Phys.  {\bf 62}  (1975)   1544

\bibitem{Hepp74}
K. Hepp, {\it The classical limit for quantum mechanical correlation functions}, Commun.
   Math. Phys. {\bf 35} (1974)  265
   
\bibitem{Litt86}
R.~G. Littlejohn,  {\it The semiclassical evolution of wave packets},  Phys.
  Rep.  {\bf 138}  (1986)   193

\bibitem{Hell81}
E.~J. Heller,  {\it Frozen Gaussians: A very simple semiclassical
  approximation},  J. Chem. Phys.  {\bf 75}  (1981)   2923

\bibitem{Hube88b}
D.~Huber and E.~J. Heller,  {\it Hybrid mechanics: A combination of classical
  and quantum mechanics},  J. Chem. Phys.  {\bf 89}  (1988)   4752

\bibitem{Hube89}
D.~Huber, S.~Ling, D.G. Imre, and E.~J. Heller,  {\it Hybrid mechanics. II},
  J. Chem. Phys.  {\bf 90}  (1989)   7317

\bibitem{Mill02}
W.~H. Miller,  {\it An alternative derivation of the Herman-Kluk (coherent
  state) semiclassical initial value representation of the time evolution
  operator},  Molecular Physics  {\bf 100}  (2002)   397

\bibitem{Bara01}
M.~Baranger, {M. A. M. de} Aguiar, F.~Keck, H.~J. Korsch, and
  B.~Schellhaa{\ss},  {\it Semiclassical approximations in phase space with
  coherent states},  J. Phys. A  {\bf 34}  (2001)   7227

\bibitem{Kong16}
X.~Kong, A.~Markmann, and V.~S. Batista,  {\it Time-sliced thawed Gaussian
  propagation method for simulations of quantum dynamics},  J. Phys. Chem. A
  {\bf 120}  (2016)   3260

\bibitem{Breu02}
H.-P. Breuer and F~Petruccione,  {\em The theory of open quantum systems},
  Oxford University Press, Oxford, 2002

\bibitem{Stru97}
W.~T. Strunz,  {\it Path integral, semiclassical and stochastic propagators for
  {M}arkovian open quantum systems},  J. Phys. A  {\bf 30}  (1997)   4053

\bibitem{Ozor09}
A.~M.~Ozorio de~Almeida, P.~de~M.~Rios, and O.~Brodier,  {\it Semiclassical
  evolution of dissipative {M}arkovian systems},  J. Phys. A  {\bf 42}  (2009)
   065306

\bibitem{Brod10b}
O.~Brodier and A.M.~Ozorio de~Almeida,  {\it Markovian evolution of {G}aussian
  states in the semiclassical limit},  Phys. Lett. A  {\bf 374}  (2010)   2315

\bibitem{Grae11}
E.~M. Graefe and R.~Schubert,  {\it Wave packet evolution in non-Hermitian
  quantum systems},  Phys. Rev. A  {\bf 83}  (2011)   060101(R)

\bibitem{Grae12}
E.~M. Graefe and R.~Schubert,  {\it Complexified coherent states and quantum
  evolution with non-Hermitian Hamiltonians},  J. Phys. A  {\bf 45}  (2012)
  244033

\bibitem{Zach05}
C.~Zachos, D.~Fairlie, and T.~Curtright,  {\em Quantum Mechanics in Phase
  Space},   World Scientific Series, Singapore, 2005

\bibitem{Ades14}
G.~Adesso, S.~Ragy, and A.~R. Lee,  {\it Continuous Variable Quantum
  Information: {G}aussian States and Beyond},  Open Syst. Inf. Dyn.  {\bf 21}
  (2014)   1440001

\bibitem{Perc98}
I.~C. Percival,  {\em Quantum State Diffusion},   Cambridge University Press,
  Cambridge, 1998

\bibitem{Grme97}
M.~Grmela and H.~C. {\"O}ttinger,  {\it Dynamics and thermodynamics of complex
  fluids. I. Development of a general formalism},  Phys. Rev. E  {\bf 56}
  (1997)   6620--6632

\bibitem{Oett97}
H.~C. {\"O}ttinger and M.~Grmela,  {\it Dynamics and thermodynamics of complex
  fluids. II. Illustrations of a general formalism},  Phys. Rev. E  {\bf 56}
  (1997)   6633--6655

\bibitem{Bowe16}
W.~P. Bowen and G.~J. Milburn,  {\em Quantum Optomechanics},   CRC Press, Boca
  Raton, FL, 2016

\bibitem{SchuValTos12}
R.~Schubert and R.~Vallejos and F.~Toscano,  {\it How do wave packets spread? {T}ime evolution on 
{E}hrenfest time scales},  J. Phys. A  {\bf 45}  (2012)
  215307 
  
\bibitem{Daley}
A. J. Daley, {\it Quantum trajectories and open many-body quantum systems}, Advances in Physics {\bf 63} (2014) 77

\bibitem{Kordas}
G. Kordas and D. Witthaut and P. Buonsante and A. Vezzani and R. Burioni and A. I. Karanikas and S. Wimberger, {\it The Dissipative Bose-Hubbard Model. Methods and Examples}, Eur. Phys. J. Special Topics {\bf 224}  (2015) 2127
  
  \bibitem{Koda15}
S.~Koda,  {\it Initial-value semiclassical propagators for the {W}igner
  phase-space representation: Formulation based on the interpretation of the
  Moyal equation as a Schr{\"o}dinger equation},  J. Chem. Phys.  {\bf 143}
  (2015)   244110

\bibitem{Mois11book}
N.~Moiseyev,  {\em Non-{Hermitian} {Quantum} {Mechanics}},   Cambridge
  University Press, Cambridge, 2011

\bibitem{Bend07}
C.~M. Bender,  {\it Making sense of non-hermitian hamiltonians},  Rep. Prog.
  Phys.  {\bf 70}  (2007)   947

\bibitem{NEFT1}
T. Arimitsu and H. Umezawa, {\it A General Formulation of Nonequilibrium Thermo Field
Dynamics}, Prog. Theo. Phys. {\bf 74} (1985) 429

\bibitem{NEFT2}
M. Am-Shallem and A. Levy and I. Schaefer and R. Kosloff, {\it Three approaches for representing Lindblad dynamics by a matrix-vector notation}, arXiv:1510.08634


\bibitem{Schl10}
M. Schlosshauer,  {\em Decoherence and the quantum-to-classical transition},
  Springer-Verlag, Berlin Heidelberg, 2010
  
  
\end{thebibliography}
\end{document}